\documentclass[aps,twocolumn,groupedaddress,superscriptaddress]{revtex4-1}
\usepackage{graphicx}
\usepackage{textcomp}
\usepackage{gensymb}
\usepackage{amsmath}
\usepackage[colorlinks = true, citecolor = magenta]{hyperref}
\usepackage{braket}
\usepackage{natbib}
\usepackage[utf8]{inputenc}
\usepackage{dcolumn,bm,mathtools}
\usepackage{float}
\usepackage{dsfont}
\usepackage{bbold}

\begin{document}

\title{Single-photon induced instabilities in a cavity electromechanical device}

\author{Tanmoy~Bera}
\email{btanmoy@iisc.ac.in} 
\affiliation{Department of Physics, Indian Institute of Science, Bangalore-560012 (India)}

\author{Mridul~Kandpal}
\affiliation{Department of Physics, Indian Institute of Science, Bangalore-560012 (India)}

\author{G.~S.~Agarwal}
\affiliation{Department of Physics, Indian Institute of Science, Bangalore-560012 (India)}
\affiliation{Institute for Quantum Science and Engineering, and Departments of 
Biological and Agricultural Engineering, and Physics and Astronomy Texas A\&M University, 
College Station, Texas 77843, USA}

\author{Vibhor~Singh}
\email{v.singh@iisc.ac.in} 
\affiliation{Department of Physics, Indian Institute of Science, Bangalore-560012 (India)}

\date{\today}

\begin{abstract}
Cavity-electromechanical systems are extensively used for sensing 
and controlling the vibrations of mechanical resonators down to their 
quantum limit. The nonlinear radiation-pressure interaction in 
these systems could result in an unstable response of the 
mechanical resonator showing features such as frequency-combs, 
period-doubling bifurcations and chaos. However, due to weak 
light-matter interaction, typically these effects appear 
at very high driving strengths.
By using polariton modes formed by a strongly coupled 
flux-tunable transmon and a microwave cavity, here we 
demonstrate an electromechanical device 
and achieve a single-photon coupling rate $\left(g_0/2\pi\right)$ 
of $160~$kHz, which is nearly 4\% of the 
mechanical frequency $\omega_m$. 
Due to large $g_0/\omega_m$ ratio, the device shows an unstable mechanical
response resulting in frequency combs in sub-single photon limit.
We systematically investigate the boundary of the unstable response 
and identify two important regimes governed by the optomechanical 
backaction and the nonlinearity of the electromagnetic mode.
Such an improvement in the single-photon coupling rate and the 
observations of microwave frequency combs at single-photon levels 
may have applications in the quantum control of the motional states 
and critical parametric sensing. Our experiments strongly suggest
the requirement of newer approaches to understand instabilities. 
\end{abstract}
	
	
\maketitle

\section*{Introduction}

Light carries momentum, and it can be used to control
and manipulate the motion of a mechanical resonator down 
the quantum regime \cite{aspelmeyer_cavity_2014}. 
Such control over the motional states is essential for the 
technological advancement as well as to probe the fundamental 
physics \cite{qiao_splitting_2023,bild_schrodinger_2023}.
In cavity-electromechanical devices, two \textit{linear} modes
namely an electromagnetic (EM) mode and a mechanical
mode are coupled with the \textit{nonlinear} radiation-pressure 
interaction
\cite{aspelmeyer_cavity_2014,clerk_hybrid_2020}.
In the microwave domain, the vibrations of 
the mechanical resonators are typically integrated 
into the EM mode using the charge modulation \cite{wollman_quantum_2015,pirkkalainen_squeezing_2015,lecocq_quantum_2015,mason_continuous_2019,peterson_ultrastrong_2019,kotler_direct_2021,wollack_quantum_2022,seis_ground_2022,das_instabilities_2023}.
Recently, cavity electromechanical devices utilizing 
the magnetic-flux modulation of Josephson inductance 
have shown interesting results such as large
electromechanical coupling rates \cite{rodrigues_coupling_2019,bera_large_2021,schmidt_sideband-resolved_2020,zoepfl_single-photon_2020}, the near 
ground state cooling of the mechanical 
resonator by four-wave-mixing \cite{bothner_four-wave-cooling_2022}, 
and by Kerr-enhanced techniques \cite{zoepfl_kerr_2023}.
Further, such devices have 
been proposed to reach the single-photon strong coupling 
regime using its linear scaling with the magnetic 
field \cite{via_strong_2015,khosla_displacemon_2018,kounalakis_flux-mediated_2020}.

Apart from providing the electromechanical coupling, 
the \textit{nonlinear} nature of the Josephson inductance 
in flux-coupled electromechanical systems can be a valuable resource \cite{bothner_four-wave-cooling_2022,zoepfl_kerr_2023}.
It allows us to control the nature of 
the electromagnetic mode to change from 
the weak-Kerr to a single photon strong-Kerr oscillator
where 
the response remains nonlinear down to 
a single-excitation.
Further, under certain control parameters, the electromagnetic 
(EM) mode can made to undergo a dissipative phase 
transition with strong fluctuations in
the photon number \cite{chen_quantum_2023}.
Therefore, coupling a mechanical resonator to such an EM mode 
using radiation-pressure interaction offers
a unique platform to study and discern the effects stemming 
from the \textit{nonlinear} nature of the radiation-pressure 
interaction \cite{arcizet_radiation-pressure_2006,das_instabilities_2023}, 
as well as the ones from the \textit{nonlinear} nature of the
electromagnetic mode \cite{chen_quantum_2023}.
Here, we demonstrate an electromechanical device consisting
of a linear cavity, a frequency tunable transmon qubit, 
and a mechanical resonator \cite{bera_large_2021}.
First, the flux-tunability of the transmon qubit 
is used to implement the electromechanical coupling.
Second, tuning of the transmon frequency in resonance with 
the cavity forms the new eigenstates due to the designed 
strong coupling.
The anharmonicity of the resultant polariton modes can 
can be controlled by transmon-cavity 
detuning \cite{bishop_nonlinear_2009,bishop_response_2010}. 
Such a tri-partite system thus allows us to control 
the electromechanical coupling and the nonlinearity of 
the EM mode in a single device. 
Therefore, by controlling the strength of the nonlinearity, 
the electromechanical effects can be explored
as the EM mode undergoes through various regimes such as 
``super-splitting", 
multi-photon transitions \cite{bishop_nonlinear_2009,shamailov_multi-photon_2010}, 
critical slowing down \cite{brookes_critical_2021}, 
photon blockade breakdown \cite{fink_observation_2017, sett_emergent_2024}, 
and transmon-ionization \cite{shillito_dynamics_2022}.

\section*{Results}
\subsection*{Device concept}
The frequency tunable transmon qubit is enabled
by a SQUID loop and a capacitor.
Fig.~\ref{fig1}a shows the schematic design of the
device. 
The coaxial cavity is placed inside a 2-axis vector 
magnet, allowing us to control the axial and the normal
components of the magnetic field independently. 
The cavity is machined from oxygen-free high conductivity 
(OHFC) copper.
The transmon qubit is fabricated on an intrinsic-Si 
chip using electron beam lithography and shadow evaporation
of aluminium. 
To achieve a larger critical in-plane magnetic 
field, we use 28~nm thin Al film to fabricate 
the device \cite{krause_magnetic_2022}.
The mechanical resonator is realized by suspending one
of the arms of the SQUID loop by selective isotropic 
dry etching of the silicon substrate.
The transmon frequency can be tuned by varying 
the magnetic-flux through the SQUID loop.
A false color scanning electron microscope image 
of the SQUID loop with the suspended mechanical 
resonator is shown in Fig.~\ref{fig1}b.
The patterned chip is placed inside 
the coaxial cavity and cooled down to 20~mK in a 
dilution refrigerator. 
The details of device fabrication, and measurement 
setup are given in Supplementary Note~1.
Since superconducting films show more resilience 
to the in-plane magnetic field \cite{krause_magnetic_2022}, 
we focus on the electromechanical coupling of the out-of-plane
vibrational mode of the mechanical resonator.
Further, by using the control over the normal component
of the magnetic field, we can cancel any out-of-plane
component of the magnetic field arising from 
the misalignment between the axial field and 
SQUID plane.

\subsection*{Measurements}
\begin{figure*}
\centering
\includegraphics[width = 165 mm]{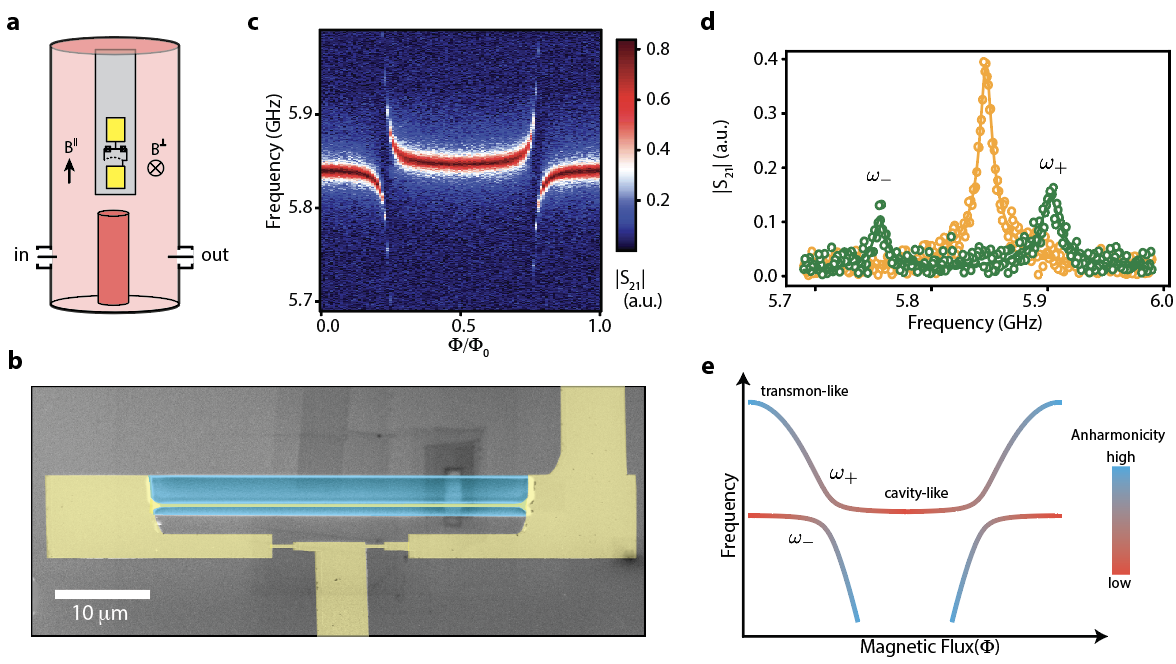}
\caption{\textbf{Concept:} 
(a) A schematic of a quarter wavelength coaxial
cavity (red) coupled to a transmon (yellow). The two 
pads provide the shunting capacitance as well as 
coupling capacitance to the cavity.
The axial and normal components of the 
magnetic-field are denoted by $B^\parallel$ 
and $B^{\perp}$, respectively.
(b) A tilted false-color scanning electron microscope 
image of the SQUID loop with embedded mechanical resonator.
The suspended Al nanomechanical resonator can be seen. 
The Al film and the silicon substrate are shown in yellow and gray 
colors, respectively. The selective-etched 
region used to suspend the mechanical resonator 
is shown by the cyan-colored region. The mechanical 
resonator has dimensions of $\text{40} \mu\text{m}\times\text{200 nm}\times\text{28 nm}$.
(c) A representative colorplot of the voltage 
transmission $|S_{21}|$ through the cavity as the 
magnetic flux through the SQUID loop is varied.
(d) The panel shows linecut from panel~(c), 
corresponding to $\Phi/\Phi_0=0.27(0.5)$ in green(yellow) 
emphasizing the vacuum-Rabi split peaks (dressed-cavity peak).
(e) A schematic to depict the change in frequency of 
the dressed modes as the magnetic flux is swept.
The gradual color shift from blue to red or vice versa 
indicates how the nonlinear character of the 
mode is changing with flux.
}
\label{fig1}
\end{figure*}

We report measurements from two similar 
devices but with substantially
different transmon-cavity coupling rates.
Detailed parameters for both the devices are listed in
Supplementary Note~2.
We first begin by measuring the voltage transmission
$|S_{21}(\omega)|$ of Device-1 through the cavity while
varying the magnetic flux through the SQUID loop.
Depending on the flux passing through the SQUID loop,
the transmon frequency can vary from
its maximum value to a minimum value.
Fig.~\ref{fig1}c shows the color plot of the cavity transmission.
As the transmon is brought in resonance with the cavity, 
the measured transmission splits into two well-separated 
dressed modes as shown in Fig.~\ref{fig1}d, demonstrating 
the strong coupling.
Provided low strength of the input power, the dressed 
cavity mode for $\Phi/\Phi_0=0.5$, and the vacuum 
Rabi-split peaks (VRS) for $\Phi/\Phi_0=0.27$ have the 
characteristic Lorentzian lineshape.
From the separation of VRS, we infer the strength of 
the transmon-cavity coupling to be 72~MHz.
The reduction in the peak height of VRS in resonant 
condition suggests that energy dissipation via 
the transmon qubit dominates over the cavity dissipation
rate.
In the resonant limit of  transmon-cavity,
it is suitable to describe the system in terms of
new eigenstates, denoted by upper and lower polariton modes 
with frequency $\omega_+$ and $\omega_-$, respectively. 
Away from the resonant limit, 
these modes exhibit a cavity-like or a transmon-like behavior 
with distinct anharmonicity as represented by 
the gradual change of color in Fig.~\ref{fig1}e.
These polariton modes remain the flux tunability 
due to underlying SQUID inductance,
allowing them to be utilized for implementing
electromechanical interaction.
Further, due to the large spectral separation between 
the two modes, we can analyze and describe
each mode's optomechanical effect separately.
Considering the upper mode, the
Hamiltonian of the tri-partite system can be reduced
to

\begin{equation}
H \simeq {\omega}_{+}\hat{a}_+^\dag\hat{a}_+
 -\frac{K_{+}}{2}(\hat{a}_+^\dag \hat{a}_+)^2
 +g_+\hat{a}_+^\dag\hat{a}_+(\hat{b} + \hat{b}^\dag) +
\omega_m b^{\dag}b ,
\label{H_eff}
\end{equation}

where $\omega_+$, $K_+$, $\omega_m$, and $g_+$ are the
upper mode frequency, anharmonicity, mechanical frequency,
and single-photon electromechanical coupling rate, respectively. 
The ladder operators of the upper polariton
mode and the mechanical mode are denoted by
$\hat{a}_+(\hat{a}_+^\dag)$ 
and $\hat{b}(\hat{b}^\dag)$, respectively. 
The single-photon electromechanical coupling rate 
of the upper polariton mode can be expressed
as $g_+\simeq {\xi}G_+B^{\parallel}lx_{zp}$,
where $G_+=d\omega_+/d\Phi$ is the flux responsivity
of the upper polariton mode, $B^{\parallel}$ is the axial 
component of the applied magnetic field, $l$ is the
length of the mechanical resonator, $x_{zp}$ is 
the quantum zero-point fluctuations of the mechanical 
resonator and $\xi$ is a geometrical factor of order 
unity and depends on the mechanical mode-shape.
It is evident that higher single-photon electromechanical 
coupling rate can be achieved by increasing the 
flux-responsivity $G_+$ and magnetic field $B^{\parallel}$.
However, the increased flux-responsivity of the dressed 
mode comes at a cost of reduced transmission.
For the experiments discussed in the next section, we 
choose an operating frequency $\omega_+$ in Device-1 that are
20 to 80~MHz detuned from the bare cavity frequency. 

\subsection*{Cavity-enabled qubit-phonon absorption}
\begin{figure*}
\centering
\includegraphics[width=145 mm]{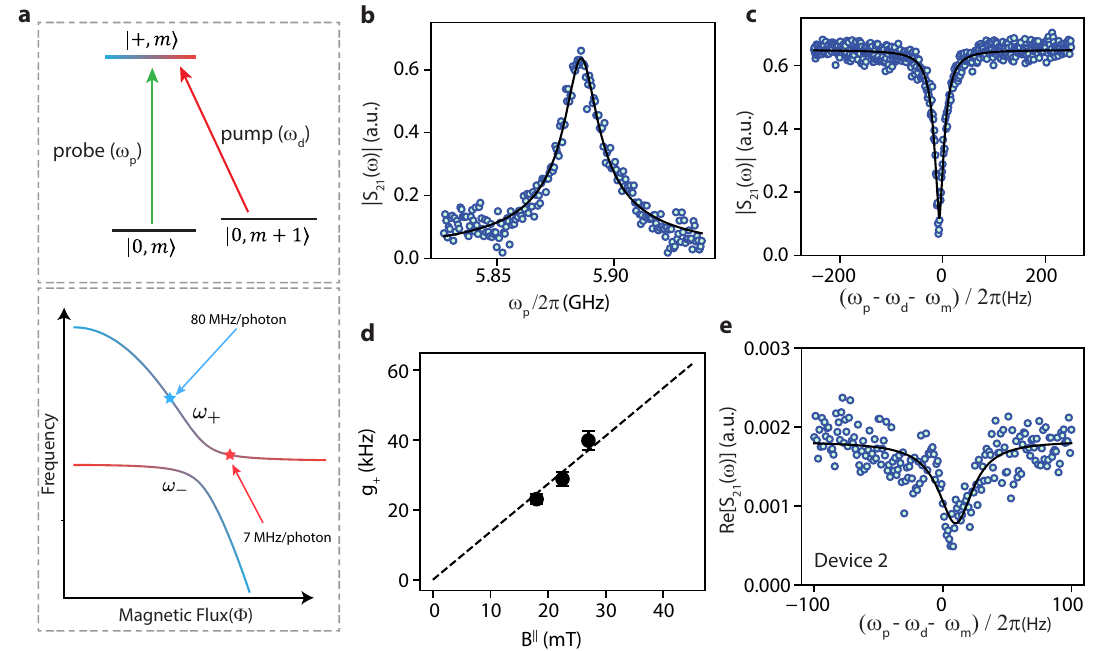}
\caption{\textbf{Cavity-enabled qubit-phonon absorption (CEQA):} 
(a) The top panel depicts the Pump and probe
scheme for the CEQA experiment. 
The states $\ket{+,m}$ denote one 
excitation in the polariton mode, along with 
$m$ excitations in the mechanical 
resonator.
The pump and probe signals are shown by the red 
and green arrows, respectively.
The lower panel schematically illustrates operating 
points for the CEQA experiment in both devices.
The red star represents the operating point
in Device-1, indicating weak nonlinearity 
in the polariton mode, whereas the blue star
represents the operating point in Device-2, 
indicating strong nonlinearity.
(b) Measurement of $|S_{21}|$ in the absence of 
the pump signal, showing the linear response of the 
polariton mode. The solid-black line is a Lorentzian fit 
to the response.
(c) In the presence of a pump signal, an absorption 
feature appears in the probe transmission. 
At $B^{\parallel}=\sim27$~mT, the polariton mode 
frequency is set to $\omega_+/2\pi\sim5.884~$GHz.
The applied pump strength is $P_i = -19~$dBm,
which corresponds to mean ``photon" occupation of 
$(5.80\pm0.07)\times10^{-2}$, and the probe signal
is 6~dB smaller than the pump.
The solid black curve is the fitted curve yielding a
single-photon coupling rate $g_+/2\pi\sim 40.0 \pm5.5~$kHz.
(d) Experimentally determined $g_+$ for three different 
values of the applied magnetic field $B_{\parallel}$.
While increasing $B^{\parallel}$, the flux-responsivity is
nominally kept constant by adjusting $B^{\perp}$.
(e) The absorption feature arising in the probe transmission
of a strongly nonlinear mode in Device-2. 
The mode frequency is chosen to be $\omega_+/2\pi\sim6.005$~GHz, which corresponds to
a qubit detuning of $\left(\omega_q - \omega_c\right)/2\pi\sim120$~MHz.
An analytical formula derived from a two-level 
system coupled to a mechanical resonator is 
used to fit the data, yielding the solid black curve.
}
\label{fig2}
\end{figure*}

To probe the electromechanical coupling, we use
a pump-probe scheme, similar to the electromagnetically 
induced transparency 
technique \cite{fleischhauer_electromagnetically_2005,weis_optomechanically_2010}.
Fig.~\ref{fig2}a shows the schematic of the different 
transitions involved in the measurement.
Two coherent signals are sent to the device, \textit{i.e.} 
a pump signal at a red detuned frequency $(\omega_+-\omega_m)$
which drives the $\ket{0,m+1}\leftrightarrow \ket{+,m}$ transition,
and a weak probe signal at $\omega_p$ (near $\omega_+$) to measure
the transmission.
We emphasize that due to the tunable nature of the polariton
modes in our device, we can carry out such experiments in both
low and strong anharmonicity limit, as shown in the lower
panel of Fig.~\ref{fig2}a.
We first perform the experiment in Device-1 with 
a ``cavity-like" mode.
Without the pump, the voltage transmission $|S_{21}(\omega_p)|$ 
shows the characteristic Lorentzian lineshape 
as shown in  Fig.~\ref{fig2}b.
In the presence of the pump, an absorption feature 
appear in the transmission spectrum, as shown 
in Fig.~\ref{fig2}c. It arises from the destructive 
absorption of the probe field from two different 
pathways \cite{agarwal_electromagnetically_2010}.
The shape of the absorption feature is determined 
by the response of the mechanical resonator and can 
be used to extract its resonant frequency and 
effective dissipation rate. 
From these measurements, we obtain the mechanical
resonant frequency and linewidth to be $\omega_m/2\pi\sim3.97$~MHz
and $\sim$13~Hz, respectively.
In addition, we determine the single-photon coupling strength $g_+$ 
by fitting the absorption feature with an analytical expression 
calculated by treating the electromagnetic mode as 
an anharmonic oscillator (Eq.~S11 of Supplementary Note~3).
We repeat the experiment for various pump strengths and 
extract the electromechanical coupling strength $g_+/2\pi\sim 40.0\pm 2.7~$kHz,
with an uncertainty arising from the statistical measurements.
Since the single-photon electromechanical coupling rate is linearly
proportional to the axial magnetic field $B^{\parallel}$, 
we carry out similar absorption experiments at different 
magnetic fields, and calculate the single photon electromechanical 
coupling rates. 
Fig.~\ref{fig2}d shows the plot of experimentally determined 
$g_+$ for different values of the magnetic field.

Next, we utilize the ``transmon-like" mode which exhibits a 
significantly larger anharmonicity compared to its dissipation rate. 
These measurements are done on Device-2, which is designed 
to have a transmon-cavity coupling of $J/2\pi\sim193~$MHz.
For the three-mode system discussed here, measurement 
of the transmon-phonon interaction is enabled by the cavity, we call
this process cavity-enabled qubit-phonon absorption (CEQA). 
In the large anharmonicity limit, the absorption feature
in probe transmission is shown in Fig.~\ref{fig2}e. 
Due to its large anharmonicity, the CEQA
feature effectively arises dominantly from the 
participation of ground and first excited states
of the  ``transmon-like" mode.
Consequently, we can approximate the electromechanical
interaction as an effective two-level system
longitudinally coupled to a mechanical resonator.
Keeping this in mind, we derive an analytical
expression for the absorption feature and
fit the experimental data, resulting in the black 
line shown in Fig.~\ref{fig2}e. Additional details pertaining to 
the calculations are given in the Supplementary Note~3.
We emphasize that due to large $g_+$ in the
system the CEQA feature appears
in both weak and strong nonlinearity regimes down
to the mean-photon occupation of $\sim10^{-1}$
and $\sim10^{-3}$, respectively. 

\subsection*{Optomechanical backaction and instability}

A high single-photon coupling rate results into significant 
optomechanical backaction at very low pump strengths.
To investigate the effects of dynamical backaction
on the mechanical resonator, we send a pump signal and 
measure the power spectral density (PSD) of the output
microwave signal while varying the pump power 
and detuning. 
Depending on the pump detuning, an imbalance between the
up- and down-scattering rates of microwave photons
by the mechanical resonator results in its cooling or heating. 
We first operate at a lower magnetic field of $B^\parallel\sim18~$mT, 
and at a mode frequency of $\omega_+/2\pi\sim5.873~$GHz, which 
corresponds to an estimated Kerr nonlinearity of $K_+/2\pi\sim~5.1$~MHz/photon
(see Supplementary Note~2).
At such operating frequency, the upper polariton mode has an estimated
flux responsivity of $G_+/2\pi\sim$~1.16~GHz$/\Phi_0$, and expected $g_+/2\pi\sim13.4\pm0.8~$kHz
(see Supplementary Note 2). 
Fig.~\ref{fig3}a shows the measurements of the effective
mechanical linewidth and the shift in the mechanical 
frequency, extracted from the PSD measurements. 
As expected, we observe a broadening of the mechanical
linewidth for negative detunings and heating and 
unstable response for the positive detunings.
With the increase in the strength of the pump signal, the 
backaction effects become enhanced, and the onset 
of unstable response shifts towards negative detuning 
due to the Kerr nonlinearity.

To theoretically understand the experimental observation,
we model the system as a weak Kerr oscillator with
energy decay rate rate $\kappa/2\pi\sim14~$MHz, estimated from
transmission spectrum. We incorporate the pump by
adding a drive term in the effective Hamiltonian
given by Eq.~\ref{H_eff}. 
To compute the backaction effects on the 
mechanical resonator, we solve the equations
of motion for the coupled modes and 
obtain the expressions of the 
effective linewidth $\Gamma_m$ and the frequency 
shift $\Delta\omega_m$ of the mechanical 
resonator \textbf(see Supplementary Note~4).
The solid black lines in Fig.~\ref{fig3}a are the 
theoretically calculated results.
We next focus on the unstable response 
of the mechanical resonator.
An unstable mechanical resonator has a large 
oscillation amplitude, and produces a frequency comb
structure in the PSD with multiple peaks
separated by $\omega_m$, as shown in Fig.~\ref{fig3}b. 
In linear cavity optomechanics, such features have 
been studied extensively, both theoretically and 
experimentally \cite{aspelmeyer_cavity_2014}.
We investigate the boundary of the unstable response
by varying the pump power and its detuning, as shown 
in Fig.~\ref{fig3}c.
This experiment is carried out on Device-2, while operating 
at $\omega_+/2\pi\sim5.82~$GHz,
corresponding to an energy decay rate of $\kappa/2\pi\sim9~$MHz
and an electromechanical coupling rate of
$g_+/2\pi\sim45.0\pm 1.9~$kHz, which is determined
from a separate CEQA experiment.
It is important to note that due to such a large coupling, the onset of
mechanical instability occurs at mean photon occupation
of 0.01 in the polariton mode.
To theoretically understand the boundary of the 
instability, we use a weak-Kerr model as mentioned
earlier (Supplementary Note~4), and
the results from these calculations are shown by the 
solid-black line in Fig.~\ref{fig3}c. 
By comparing the two theoretical results, we observe that
due to the nonlinearity present in the system, the onset
of instability shift toward lower frequencies as the pump
power is increased. 

\subsection*{Mechanical Instabilities in single-photon strong Kerr limit}

We first note that the response of the circuit-QED system
can be substantially different at high driving powers.
Fig.~\ref{fig4}a and b show the cavity transmission $|S_{21}|$ of Device-2 while varying $B^\perp$.
Apart from the dominating vacuum-Rabi splitting,
additional transitions arising from higher levels can also be seen.
A schematic of the uncoupled and new (polariton) eigenstates is shown in 
the right panel of Fig.~\ref{fig4}c. The left panel depicts
the uncoupled states as $\ket{n_c,n_q}$, where $n_c$ 
and $n_q$ are the number of excitation in cavity
and transmon, respectively.
The symmetric and anti-symmetric combination of single
excitation states $\ket{1,0}$ and $\ket{0,1}$ are denoted by $\ket{+}$ and $\ket{-}$, respectively.
Similarly, the new eigenstates in the two-excitation
manifold are labeled as $\ket{\alpha}$,$\ket{\beta}$ and $\ket{\gamma}$.
The additional peaks in the transmission spectrum of
Fig.~\ref{fig4}b are arising from the higher-level transitions, namely
$\{\ket{+}, \ket{-}\} \leftrightarrow \{\ket{\alpha},
\ket{\beta}, \ket{\gamma}\}$.
Such transitions are possible with a single frequency
drive due to non-zero thermal occupation of $\ket{+}$ 
and $\ket{-}$ states.
To identify these transitions,
we treat the transmon as a nonlinear oscillator 
and model the system using an
extended Jaynes-Cummings Hamiltonian added with a
coherent drive. We then numerically solve the 
Markovian master equation to compute 
the voltage transmission. The results are plotted     
as solid black line in Fig.~\ref{fig4}b.
It can be seen that the transition frequencies 
corresponding to the peaks in the transmission spectrum
match closely with the numerical results.
Clearly, these transitions arise from the single photon 
excitation and are distinct from the multi-photon 
transitions \cite{bishop_nonlinear_2009}, which
usually occur at stronger drive strengths.
We next activate the electromechanical coupling by applying 
the parallel magnetic field $B^{\parallel}\sim 9~$mT.
and investigate the boundary of the unstable response of 
the mechanical resonator in a wider span of pump power 
and frequency.
Fig.~\ref{fig5}a and c show the region of unstable response from 
Device-2 at two different values of flux-operating point,
and Fig.~\ref{fig5}b and d show the corresponding transmission spectra 
at $B^{\parallel} = 0$.
For the operating flux in Fig.~\ref{fig5}a,
the transmon qubit becomes near resonant to the cavity,
and the electromechanical coupling rate of the
higher frequency polariton mode
is estimated to reach $g_+/2\pi\sim 160~$kHz,
which is nearly 4\% of the mechanical mode 
frequency $\omega_m$.
As a consequence, the onset of mechanical
instability takes place at a very low mean photon 
occupation of $n_d\sim3\times 10^{-4}$. 
The onset of mechanical instability coincides
with the blue-detuned region of the transition frequency
$\omega_+$ and $\omega_-$, similar to the situation in the linear cavity 
optomechanics.
As pump power is increased, the mechanical instability
emerges near the higher transition frequencies
of $\omega_{-\alpha}$ and $\omega_{-\beta}$.
With further increase in the pump power, 
we observe instabilities near the bare cavity frequency.
Surprisingly, we do not see any instability branch 
surrounding the transition frequency $\omega_{+\gamma}$. 
We also note that unstable mechanical response 
resulting into frequency comb features in PSD 
persist even beyond $-5~$dBm pump power.
At these powers, we expect transmon to be ``ionized", and SQUID
would operate in non-zero voltage regime. 
The cavity transmission shows the bare-cavity response as
shown in the lower panel of Fig.~\ref{fig5}b.
Fig.~\ref{fig5}c shows similar measurements 
but at a different flux-operating point, such 
that the transmon qubit is approximately $240$~MHz
detuned above the cavity frequency. 
An instability region corresponding to single-photon 
transitions with frequency $\omega_{-\beta}$ and
$\omega_{+\gamma}$ emerges, along with the instability
branches surrounding the frequencies $\omega_+$ and $\omega_-$.
At moderately higher power, the instability region near
the two-photon transition corresponding to frequency
$\omega_\gamma/2$ can be seen. 
We also observe the super-splitting of the $\omega_-$ peak, 
resulting in an abrupt widening of the instability region. 

\subsection*{Semiclassical analysis and a model based on quantum two-level system}

To gain insight into these observations, we  plot
the transmission measurements $|S_{21}|$ over 
a large frequency at $B^{\parallel}=0$ and the boundary of 
instability, together as shown in Fig.~\ref{Fig6}a.
The instability parameter-space can be divided 
into three regions -- (i) the low-power region, where 
the single-photon instabilities 
stem from the lower transitions, 
(ii) the mid-power region, where higher energy single-
and multi-photon transitions, super-splitting, 
bistability of the electromagnetic mode are important, and 
(iii) the ``ionization" region, where the frequency-comb 
persists despite the SQUID operating in the non-zero 
voltage regime.
As the EM mode emerges from the strong coupling between the 
transmon and the cavity, we model it by expanding it using polariton 
eigenstates to capture the low-power behavior. 
Such an approach effectively breaks down the EM mode into 
independent two-level systems, which are longitudinally 
coupled to a mechanical resonator. 
Details of the model are included in the 
Supplementary Note~6. 
Results from these calculations are plotted in 
Fig.~\ref{Fig6}b. As expected, the theoretical model 
is able to capture the behavior at low pump powers, 
however, it does \textit{not} capture the experimental observations 
at intermediate or high pump powers.
For intermediate powers, we attempt a model based on the semi-classical 
analysis while treating the transmon as an anharmonic oscillator (details 
are included in Supplementary Note~5).
We first note that a driven transmon-cavity system alone (no optomechanical 
coupling) can have bistable and unstable solutions of the intra-cavity 
field \cite{khan_frequency_2018}.
We use semi-classical approach and perform the linear stability 
test \cite{strogatz_nonlinear_2019}. We identify the regions with one unstable fixed-point (FP), 
and regions with one unstable and two stable FP, as shown in Fig.~\ref{Fig6}c.
In these regions, the ``intra-cavity" field is expected to show either an 
unstable or a bistable behavior resulting in large photon-fluctuations.

Next, we include the optomechanical interaction, and
obtain the fixed-points.
The regions of the mechanical instability is shown by the dashed
lines in Fig.~\ref{Fig6}c.
At low powers, the semi-classical analysis could produce the 
mechanical instability region near the $\omega_{\pm}$ polaritons, arising from
the optomechanical backaction.
In addition, we note that in the optically bistable region, one of the 
stable solutions also give rise to mechanical 
instability.

Clearly, the quantum model based on the polariton-basis, and 
the three-mode semi-classical analysis do not capture
the experimentally observed behavior. While the model based
on the polariton-eigenstates captures the low-power behavior, it
is not as effective in mid-power range.
Particularly, in the middle part of the region (ii), the transmon-cavity 
system undergoes a dissipative phase transition and can take a 
long time to reach the equilibrium \cite{brookes_critical_2021,chen_quantum_2023,sett_emergent_2024}. 
In this region, the presence of unstable response in the experimental data, 
and the absence of it in the modelled results strongly suggest the role of 
fluctuating photon-pressure on the mechanical resonator during the transition. 
These observations would require theoretical investigations beyond 
the standard semi-classical or two-level approach \cite{minganti_spectral_2018}.

\section*{Discussion}

To summarize our findings, our current experiment 
reaches a single-photon coupling rate which is nearly 4\% of
the mechanical resonator frequency. It highlights that 
instabilities arising from the residual thermal occupations
of the mechanical and EM modes would become important
as one reaches the single-photon strong coupling 
regime $g_0\gtrsim\{\kappa,\omega_m\}$ unless the mechanical oscillator 
is cooled to its quantum ground state. 
Single-photon strong coupling regime seem experimentally
feasible as transmon qubits have been 
shown to operate in higher magnetic 
fields \cite{krause_magnetic_2022}.

With such promising futuristic parameters, it is feasible
to achieve ground state cooling via sideband driving 
below the level of a single photon, and it further suggests 
ways to prepare non-classical mechanical states 
including Schrodinger-cat \cite{li_generation_2018,khosla_displacemon_2018,kounalakis_flux-mediated_2020}.
Such methods and techniques can be extended to the low frequency 
flux-family superconducting qubits to realize transverse electromechanical 
couplings, and thus extending the toolbox available with the flux-coupled devices 
\cite{najera-santos_high-sensitivity_2024}. 
The generation of microwave frequency combs at single-photon 
level could have applications in quantum 
sensing \cite{di_candia_critical_2023,tang_enhancement_2023}.
Our experimental observations further demand theoretical investigation 
into the parametric instabilities in single photon limit.

\section*{Materials and methods}

The coaxial cavity is machined from oxygen-free high 
conductivity copper. The central post (solid cylinder 
in the lower half) has a length of 11.5~mm and diameter 
of 2.5~mm. The inner diameter of the outer cylinder 
is 5.5~mm, and the total height of the cavity is 20.5~mm.
The transmon qubit is fabricated on an 
intrinsic silicon-(100) substrate.
Using EBL and shadow evaporation of aluminum, we 
fabricate the device using a single step of lithography. 
The evaporated aluminum film is annealed under ambient 
conditions to transform the compressive stress to tensile
stress. 
A selective etching of silicon is then carried out
to release the nanowire (mechanical oscillator) 
from the substrate.
To nullify the effects of annealing and etching on
Josephson junction inductance, the oxidation parameter
is tuned accordingly at the time of deposition so that 
we get the desired junction resistance at the end of
all processes.
Finally, the silicon chip is mounted inside the coaxial 
cavity and the cavity is placed inside a home-built 
vector magnet setup. This assembly is placed inside a 
double layer of infrared and magnetic field shields, 
and mounted to the mixing chamber plate for cooling
down to 20~mK for measurements.

\section*{Acknowledgments}      
T.B. thanks Bhoomika Bhat, Harsh Vardhan Upadhyay 
and S.~Majumder for assisting during the device 
fabrication and helpful discussions.
G.S.A. thanks the Infosys Foundation Chair of the Department of
Physics, IISc Bangalore, which made this collaboration
successful.
This work is supported by the Air Force Office
of Scientific Research under
Award No. FA2386-20-1-4003.  The authors acknowledge the support 
under the CoE-QT by MEITY and QuEST program by DST, Govt. of India. 
The authors acknowledge device fabrication facilities at CeNSE, 
IISc Bangalore, and central facilities at the Department of 
Physics funded by DST (Govt. of India).

\section*{Author contribution}
V.S. conceived and supervised the experiment. T.B. fabricated 
the device and performed the measurements. T.B. and V.S. have done
the data analysis. T.B., M.K., and G.S.A. carried out the 
theoretical calculations. All the authors have contributed 
in preparing the manuscript.

\section*{Competing interests}
The authors declare no competing interests.

%

\begin{figure*}
\centering
\includegraphics[width=145 mm]{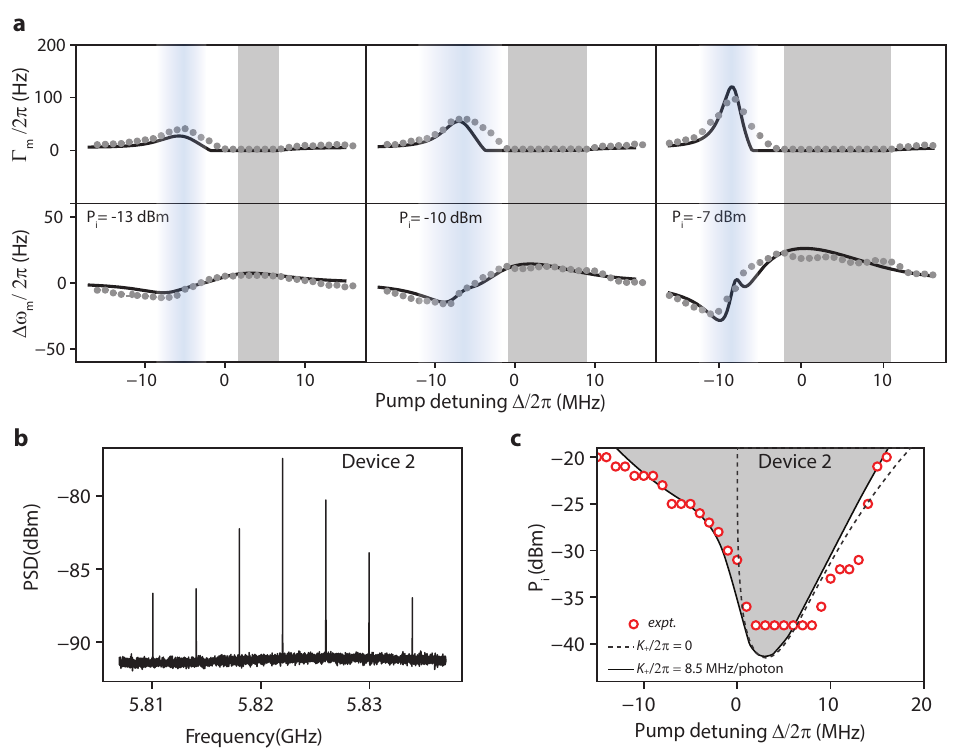}
\caption{\textbf{Dynamical backaction on the mechanical resonator:} 
(a) Measurements from Device-1.
Total mechanical linewidth and change in the mechanical 
frequency  extracted from the power spectral density (PSD) 
measurements as the pump 
detuning is varied at a constant pump power. 
From left to right panels, the pump power is increased. 
For these measurements, the dressed mode is tuned to 
$\omega_+/2\pi\sim5.873$~GHz and the axial magnetic field 
is set to $B^\parallel\sim18~$mT, resulting in an electromechanical
coupling rate of $g_+/2\pi\sim13.4\pm 0.8$~kHz.
The solid-black lines are the results of the theoretical 
calculations based on a Kerr-like oscillator model.
The blue-color gradient and gray-shaded region denote the 
cooling and unstable response of the mechanical resonator.
(b) PSD of the output microwave signal showing the formation
of the frequency combs when the mechanical 
resonator becomes unstable. The peaks are separated by 
$\omega_m$. The measurement is performed using 
a resolution bandwidth of 5~kHz.
(c) The boundary of the mechanical instability in the parameter
space of the pump detuning and power.
These measurements are carried out on Device-2 with polariton
mode frequency $\omega_+/2\pi\sim5.82$~GHz, and  applied
magnetic field $B^{\parallel}\sim 9~$mT , which resulted in 
$g_+\sim 45\pm 2.7$~kHz.
The dashed line is the theoretical prediction
based on a linear oscillator model ( zero Kerr nonlinearity)
with the same electromechanical coupling.
}
\label{fig3}
\end{figure*}

\begin{figure*}
\centering
\includegraphics[width=165 mm]{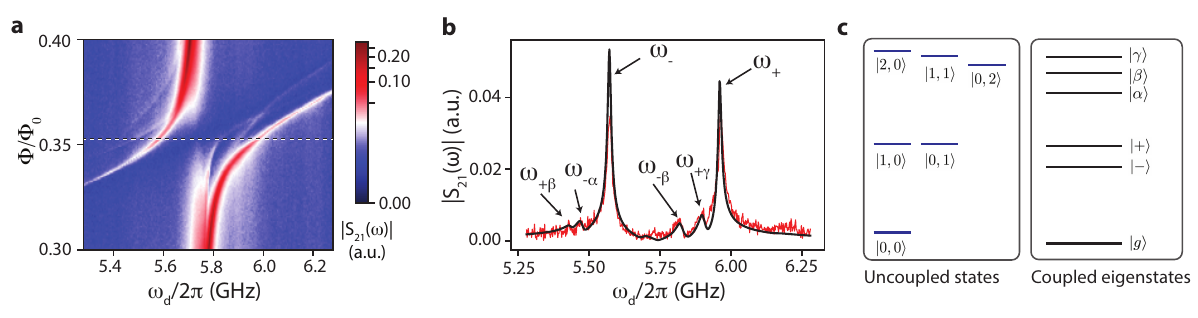}
\caption{\textbf{Higher-energy single-photon transitions of  transmon-cavity system:}
(a) Colorplot showing the voltage transmission 
$|S_{21}(\omega)|$ through the cavity for Device-2 at relatively 
stronger probe. To suppress
effects arising from the electromechanical interaction, no $B^{\parallel}$ is applied.
(b) A linecut from $|S_{21}|$ measurement shown in panel-(a)
at the position shown by the dashed line. 
A small $B^{\perp}$ is used to tune transmon frequency 
while $B^{\parallel}$ is kept at zero.
The solid line is the result from numerical calculations, 
while considering a small thermal population in both
transmon and cavity modes. Peaks are marked 
according to the transition frequencies.
(c) Left panel shows the energy levels of uncoupled
states of the transmon-cavity system. The right panel depicts 
energy-eigenstates resulting from the strong coupling.
}
\label{fig4}
\end{figure*}

\begin{figure*}
\centering
\includegraphics[width=165 mm]{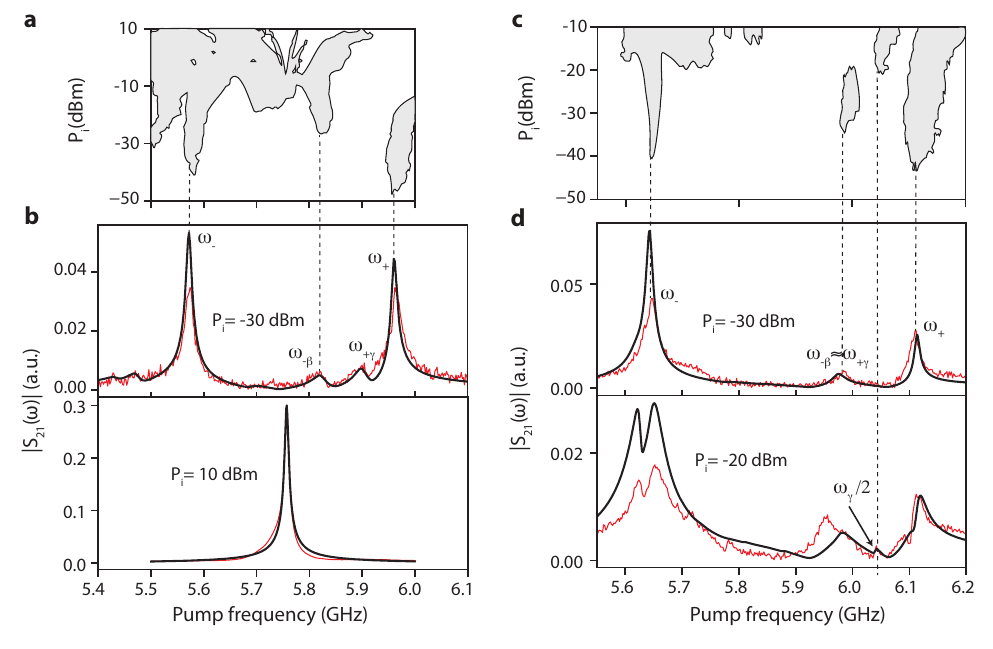}
\caption{\textbf{Mechanical instabilities due to single-, multi-photon 
transitions and super-splitted vacuum Rabi mode:}
(a, c) show the regions of the mechanical instability 
in gray color as pump power $P_i$ and frequency 
$\omega_d$ are varied. 
The measurements are carried out on Device-2 
at two different flux bias points corresponding to 
$\left(\omega_q - \omega_c\right)/2\pi\sim40$~MHz, 
and 240~MHz.
(b, d) show the transmission of the cavity at two different
drive powers  at $B^{\parallel}=0$~mT .
The black lines represent the results of the numerical
calculations that took into consideration transmon qubit's 
anharmonicity and treat it as a Kerr oscillator.
The numerical calculations capture the transition 
frequencies, however it does not capture 
the shape accurately presumably due power dependent 
relaxation and dephasing rates of transmon. 
The dotted-lines are drawn as a guide for the eye to connect
to the instability onset points in panel (a,c). 
}
\label{fig5}
\end{figure*}

\begin{figure*}
\centering
\includegraphics[width=165 mm]{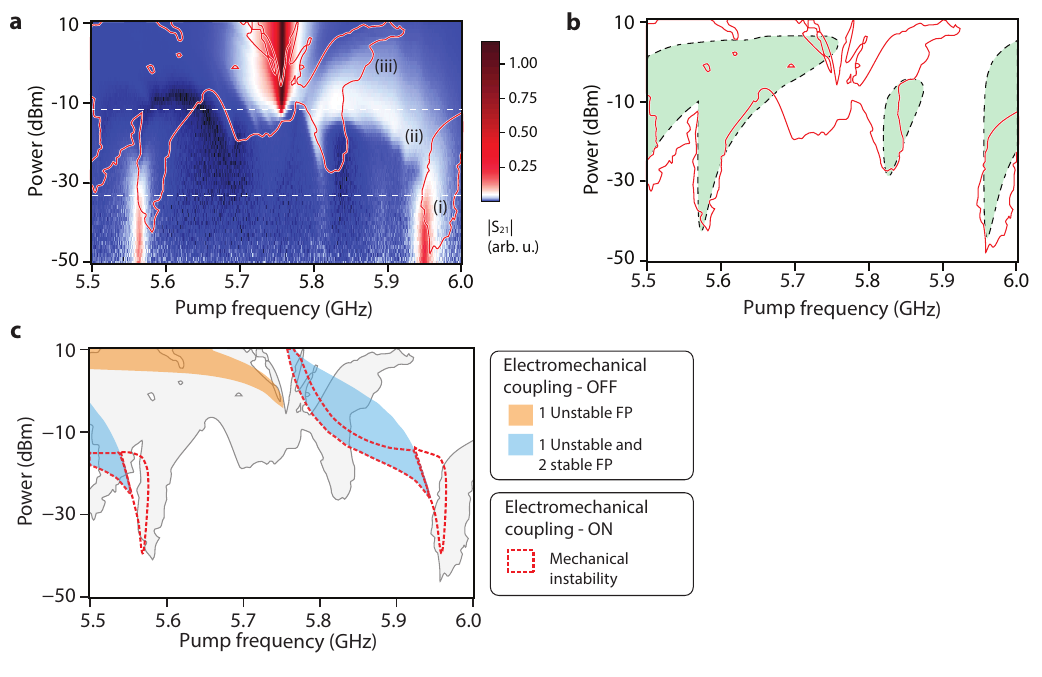}
\caption{\textbf{Insights into the instability region:}
(a) Transmission $|S_{21}(\omega)|$ of Device-2 as the pump 
power is varied. For a direct comparison with the various
transitions relevant at high powers, the boundary of 
instability region (as shown in Fig.~\ref{fig5}(a)) is 
plotted on the top (red curve).
(b) The region with light green color shows
the mechanical instability
computed by modeling the electromagnetic mode
using polariton eigenstates, while considering
their individual coupling to the mechanical mode with
varying strengths. For comparison, the boundary 
of the instability is replotted (red curve).
Panel (c) summarizes the results from semi-classical
analysis of the system. It shows different regions with one 
or more stable and unstable FP when
the optomechanical coupling is set to zero.
Region of mechanical instability due to optomechanical 
backaction for the stable FPs of the system is also 
included. The experimental boundary of mechanical instability
is plotted in light-gray color.
}
\label{Fig6}
\end{figure*}

\clearpage

\setcounter{figure}{0}
\renewcommand{\thefigure}{S\arabic{figure}}
\renewcommand{\thetable}{S\arabic{table}}
\renewcommand{\theequation}{S\arabic{equation}}
\renewcommand{\thesection}{SUPPLEMENTARY NOTE~\arabic{section}}




\section{}
\subsection{Device Fabrication and experimental setup}

The devices are fabricated on a cleaned 2.5~mm$\times$7~mm 
silicon-(100) substrate. A single-step electron beam lithography 
(EBL) process is used to pattern the substrate with 
a bi-layer resist stack of MMA-EL-11 and PMMA-950-A4. 
Subsequently, aluminum(Al) films are deposited using 
shadow evaporation technique with an intermediate \textit{in-situ} 
oxidation step. 
We found the evaporated films to be under compressive stress 
after the deposition, which is not quite suitable for the 
release of the Al-resonator. To convert it to the tensile stress,
the chip is annealed at 180$\degree$C for 15 min in the 
ambient environment. It leads to a change in the tunnel junction
resistance of the SQUID at room temperature, as mentioned 
in Table~\ref{table-s1}.

\begin{figure}
\centering
\includegraphics[width=85 mm]{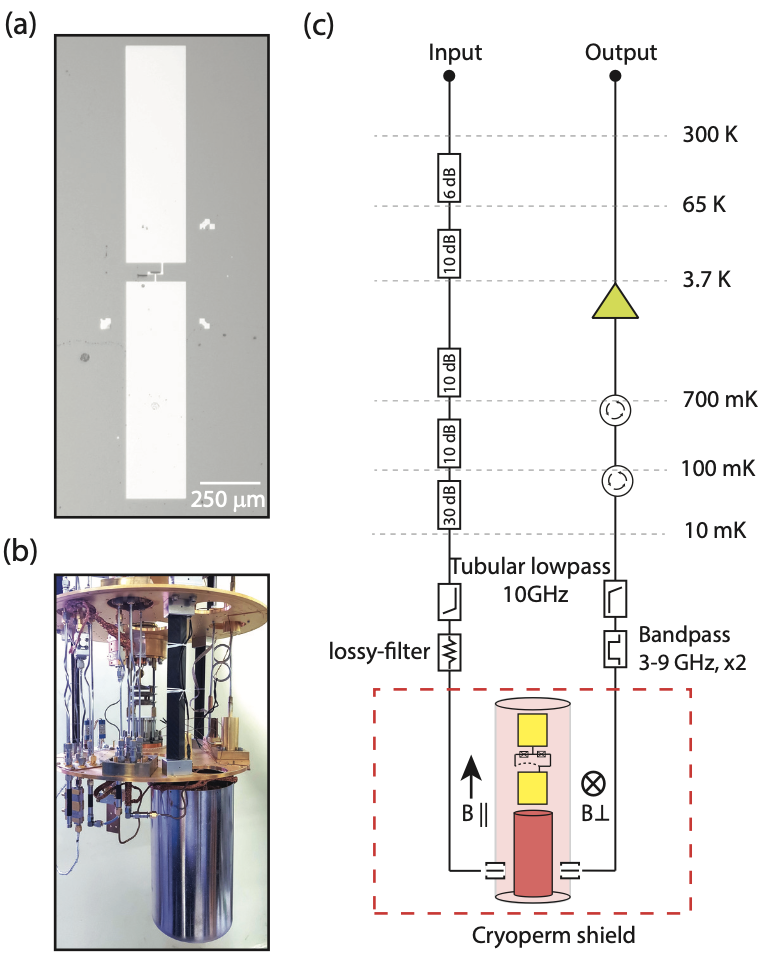}
\caption{\textbf{Device details and the measurement setup:} 
(a) An optical microscope image of the qubit device patterned
on a silicon substrate. The white rectangles are the qubit 
antenna pads. In between the pads, a SQUID loop with a suspended 
nanowire is patterned.
(b) Entire device assembly, inside a two-layer shield,
attached to the mixing chamber plate of the dilution 
refrigerator is shown.
(c) Schematic of the measurement setup, showing input
and output lines with the attenuation, amplifier, and 
filters.
}
\label{Msmt setup}
\end{figure}

Next, we carry out electron beam lithography using a single layer of 
PMMA resist and pattern a rectangular window surrounding the nanowire. 
It is followed by a reactive ion etching (RIE) process, where the silicon 
underneath the wire is etched out. The etching process is done
in two steps, using SF$_6$ gas only. 
In the first step, silicon is anisotropically etched by using
a low process pressure ($\approx9$~mTorr). It is then followed by 
an isotropic etch at higher process pressure ($\approx95$~mTorr). 
The isotropic etch step removes silicon underneath the 
nanobeam and makes it suspended.
Without breaking the vacuum, PMMA ashing is carried out to 
remove any residual resist on the substrate. Fig.~\ref{Msmt setup}(a) 
shows the optical image of the qubit fabricated on the silicon chip.
The etching process further affects the tunnel resistance 
of the junctions. We have consistently seen a reduction in 
the tunnel resistance by $40-45$\% while 
annealing the substrate and an increase in the 
resistance by $15-20$\% after the etching
process.
To accomodate these changes, the oxidation parameters 
during the junction fabrication are tuned to get the 
target junction resistance after the final step.
Finally, the chip is placed inside a coaxial cavity, and
then inside a home-built vector magnet setup. 
We use two layers of concentric shielding cans to protect 
the device from the infrared radiation and stray magnetic 
field. The radiation tight inner can is coated with an 
IR absorbing layer, and the outer can is made of cryo-perm, 
which helps in reducing the magnetic field fluctuations 
at the sample.
The entire assembly, mounted to the mixing chamber plate 
of dilution refrigerator is shown in Fig.~\ref{Msmt setup}(b).
Fig.~\ref{Msmt setup}(c) shows the schematic of the 
complete measurement setup used in the experiment.

\subsection{Derivation of the single-photon coupling rate}
The single-photon electromechanical coupling rate 
is defined as the change in resonance frequency due to
the zero point fluctuation of the mechanical resonator.
For the upper polariton mode, it can be written as
\begin{equation}
    g_+ = \frac{d\omega_+}{dx}x_{zpf} =
    \frac{d\omega_+}{d\Phi} \frac{d\Phi}{dx}x_{zpf}
    = G_+ \frac{d\Phi}{dx}x_{zpf} \text{,}
    \label{g_0}
\end{equation}
where $G_+= d\omega_+/dx$ is the flux responsivity, and $\Phi$ is the 
total magnetic flux passing through the SQUID loop.
The magnetic-flux $\Phi$ through the SQUID loop can be written as
$$
\Phi= \Vec{B}.\Vec{A}= B_\perp l (w + a.x_\parallel)
+ B_\parallel l ax_\perp .
$$
The first-term comes due to the out-of-plane magnetic field. 
It has a static component and a component arising from 
in-plane motion $x_\parallel$ of the suspended nanowire.
The second term originates from to the out-of-plane motion 
of the beam $x_\perp$ and the in-plane magnetic filed.
Upon substitution in Eq.~\ref{g_0}, we get
\begin{equation}
    g_+= B_\perp lx_{zpf} + B_\parallel lx_{zpf}.
\end{equation}
Here we have assumed that in-plane and out-of-plane 
mode of vibrations are nearly degenrate and therefore
results in the same vacuum zero-point fluctuations $x_{zpf}$. 
Since $B_\parallel \gg B_\perp$, the first term can be 
ignored.

\section{}
\subsection{Device Parameter tables}
\begin{table}[h]
	\begin{tabular}{|p{50mm}|p{15mm}|p{20mm}|}
		\hline
		\textbf{Device-1}& \textbf{Symbol}& \textbf{Value}  \\ \hline
        Cavity frequency  & $\omega_c/2\pi$  & 5.846~GHz   \\ \hline
        Bare cavity decay rate  & $\kappa_{b}$  & 8~MHz   \\ \hline
		Maximum qubit frequency  & $\omega_q^0/2\pi$  & 7.38~GHz  \\ \hline
		Qubit-cavity coupling rate & $J/2\pi$ & $72~$MHz \\ \hline
		Measured transmon anharmonicity  & $-\alpha_T/2\pi$  & $-284~$MHz  \\ \hline
		Tunnel resistance after deposition & $R_n$  & 8.9~k$\Omega$\\ \hline
        Tunnel resistance after annealing & $R_n$  & 5~k$\Omega$\\ \hline
        Tunnel resistance after etching & $R_n$  & 5.9~k$\Omega$\\ \hline
		Josephson inductance of SQUID & $L_J$ & 7~nH \\ \hline
		Mechanical resonator length & $l$ &  $\sim40~\mu$m \\ \hline
        Mechanical resonator width & $b$ & $\sim$250~nm\\ \hline
		Mechanical resonator thickness & $d$ & $\sim$28~nm \\ \hline
		Mass of the mechanical resonator & $m$ & $\sim$0.75~pg \\ \hline
		Mechanical resonator frequency  & $\omega_m/2\pi$ & $\sim$3.97~MHz\\ \hline
		Maximum axial magnetic field & $B_{max}$ & $\sim45~$mT \\ \hline	
            Product of input-line attenuation and 
            input coupling rate & $A/\kappa_{in}$ & $\sim17444$~s \\ \hline
	\end{tabular}
	\caption{Summary of the key parameters of the first sample studied.}
	\label{table-s1} 
\end{table}

\begin{table}[h]
	\begin{tabular}{|p{50mm}|p{15mm}|p{20mm}|}
		\hline
		\textbf{Device-2}& \textbf{Symbol}& \textbf{Value}  \\ \hline
        Bare cavity frequency  & $\omega_c/2\pi$  & 5.744~GHz   \\ \hline
        Bare cavity decay rate  & $\kappa_{b}$  & 8~MHz   \\ \hline
		Maximum qubit frequency  & $\omega_q^0/2\pi$  & 8.26~GHz  \\ \hline
		Qubit-cavity coupling rate & $J/2\pi$ & $193~$MHz \\ \hline
		Measured transmon anharmonicity  & $-\alpha_T/2\pi$  & $-300~$MHz  \\ \hline
		Mechanical resonator length & $l$ &  $\sim40~\mu$m \\ \hline
            Mechanical resonator width & $b$ & $\sim$250~nm\\ \hline
		Mechanical resonator thickness & $d$ & $\sim$28~nm \\ \hline
		Mass of the mechanical resonator & $m$ & $\sim$0.75~pg \\ \hline
		Mechanical resonator frequency  & $\omega_m/2\pi$ & $\sim$3.97~MHz\\ \hline
		Maximum axial magnetic field & $B_{max}$ & $\sim9~$mT \\ \hline		
            Product of input-line attenuation and
            input coupling rate & $A/\kappa_{in}$ & $\sim1647$~s \\ \hline
	\end{tabular}
	\caption{Summary of the key parameters of the second sample studied.}
	\label{table-s2} 
\end{table}

Table~\ref{table-s1} and \ref{table-s2} list the parameters
of the devices used in the experiment.

\subsection{Calibration of input-line attenuation}

\begin{figure*}
\centering
\includegraphics[width= 165 mm]{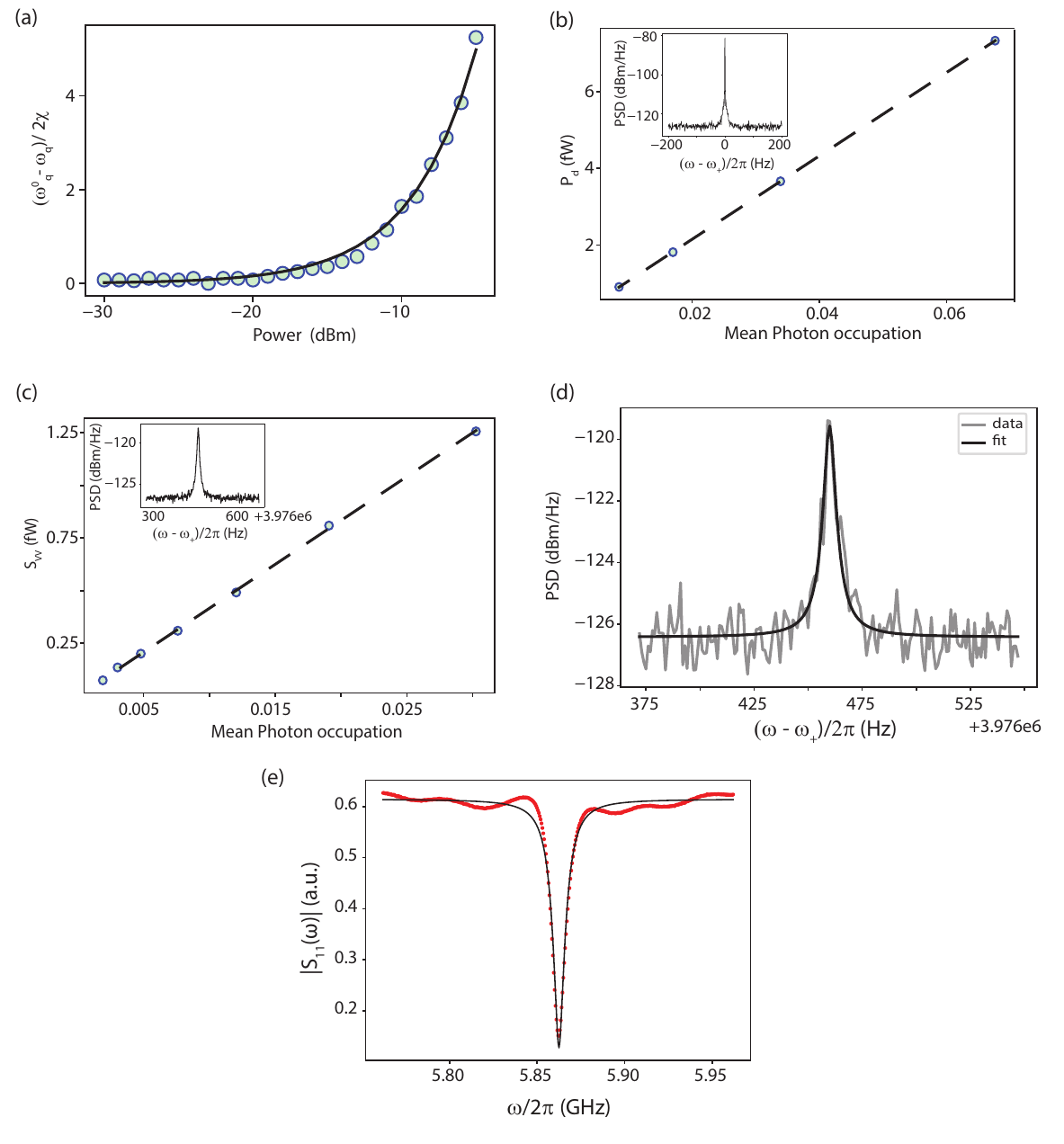}
\caption{All results are from Device-1. (a) Plot of the experimentally determined mean
photon occupation with the input probe power. The data
points show the normalized shift in the qubit frequency while
performing \textit{ac}-Stark shift measurement.
The solid black line is the fit that relates the 
injected microwave power to the mean photon occupation 
of the cavity. 
(b) Plot of the transmitted power $P_d$ at 
$\omega_+$ with mean dressed-mode occupation $n_d$. 
The dashed line is the linear fit and its slope
is used to estimate the net output gain.
This measurement is carried out at a
mode frequency of  5.864 GHz and 
applied magnetic field of $B^{\parallel}\sim18$~mT. 
The inset shows the output signal, recorded using a spectrum
analyzer, when a pump signal at $\omega_+$ is injected 
into the device.
(c) The output microwave power at the sideband peak 
$S_{VV}$ is plotted with mean dressed mode
occupation. The dashed line is a linear fit, and its
slope is used to estimate mean thermal phonon occupation.
Inset of (c) shows the mechanical resonator's
spectrum. It is the power spectral density (PSD) of the 
lower mechanical sideband for a pump at $\omega_+$.
(d) PSD of the mechanical resonator when axial
magnetic field is 18~mT, and dressed mode frequency is
5.864 GHz. From the Lorentzian fit (black-line) to the data,
we estimate the mechanical resonator frequency $\omega_m/2\pi\sim3.97~$MHz 
and intrinsic linewidth $\gamma_m/2\pi\sim6~$Hz.
(e) Reflection measurement from the output coupling port
at a temperature of 1~K. The black line is from the fitted model.
}
\label{calib}
\end{figure*}

To estimate the total attenuation in the input line, 
we use the ac-stark shift measurement. We tune the dressed
transmon frequency of Device-1 to $\omega_q/2\pi\sim5.325~$GHz 
where it couples dispersively to the cavity. 
Using the two-tone spectroscopy 
technique, we measure the transmon qubit spectrum
while probe power is varied.
With the increase in probe power, the qubit transition
frequency shifts, and it is given by 
$\omega_q\prime = \omega_{q} - 2 n_d \chi$, where $n_d$ 
and $\chi$ are mean intracavity probe photon occupation 
and the dispersive shift of the qubit, respectively. 
The dispersive shift is given by
$\chi= -\frac{J^2}{\Delta}
\frac{\alpha_T}{\Delta-\alpha_T}$, where $\Delta=\omega_q -\omega_c$ 
is the detuning between qubit and cavity.
In a separate measurement, we estimate the dispersive shift 
$ -3.5\pm0.126~$MHz of the qubit.
The experimentally computed intracavity photon
$(\omega_q^\prime -\omega_q)/2\chi$ 
is plotted with the input probe power in Fig.~\ref{calib}(a). 
Thus, it allows us to estimate the product of 
the total input line attenuation and the coupling rate 
of the input port for Device-1.
The same procedure was carried out for
Device-2 as well.
The estimated attenuation for both devices is given in 
Table~\ref{table-s1} and \ref{table-s2}.
This parameter allows us to calculate the mean photon
occupancy in a mode for a specific pump power and
energy decay rate of the mode.

\subsection{Calibration of the net output gain}
To calibrate the net output gain, we send a pump 
signal at frequency $\omega_+$ and record the 
transmitted power $P_d$ at the same frequency.
The Inset of Fig.~\ref{calib}(b) shows the measured 
power spectral density (PSD) recorded using a spectrum 
analyzer.
The transmitted power at pump frequency is given 
by $P_d/\hbar\omega_+ = A_P n_d \kappa_e$,
where $A_P$ is the net output power gain, 
$n_d$ is the mean occupation of the EM mode
due to the coherent pump, and $\kappa_e$ is the 
coupling rate of the output port. 
We vary the pump signal strength and measure
$P_d$ in a spectrum analyzer. Using the
input line attenuation, given in the device
parameter table, and the dressed 
mode decay rate of $\kappa/2\pi\sim9.7\pm 0.1~$MHz, we
can estimate the mean photon occupation $n_d$
for all pump powers.
The measured $P_d$ is then plotted against the 
mean dressed mode occupation $n_d$, as depicted 
in Fig.~\ref{calib}(b).
From the slope of the linear fit, we estimate the net 
gain. 
Using the output coupling rate of 
$\kappa_e/2\pi \approx 6.2$ MHz (discussed in 
the following section) the net gain of the output
line is estimated to be $A_P = 58.5$ dB.
The same exercise is carried out in
Device-2 as well, resulting in a net gain
of $A_P = 64.3$~dB. 
The net gain of the output line can be used 
to estimate mean-photon occupancy $n_d$ of the mode, using
 $n_d = P_d/(A_P\kappa_e\hbar\omega)$.
The reported mean-photon occupation for the experimental
result shown in Fig.5(a) of the main text is 
determined from this method.

\subsection{Estimation of the effective mechanical mode temperature}
A pump signal, tuned to the dressed mode frequency
$\omega_+$ produces two sidebands at $\omega = \omega_+
\pm~\omega_m$ due to the thermal motion of the mechanical 
resonator.
The output microwave power spectral density (PSD) of the sideband 
peak is given by,
$$ \frac{S_{VV}}{\hbar\omega}= A_P\left(\frac{1}{2} + n_{add} +
 \frac{k_e}{\gamma_m}\frac{16g_+^2 n_d n_m} {(k^2+4\omega_m^2)}\right).$$

We pump the dressed mode at zero detuning and 
record the PSD of the lower mechanical sideband using 
a spectrum analyzer.
In the inset of Fig.~\ref{calib}(c), we show a 
representative measurement 
of the microwave PSD showing the mechanical mode.
Fig.~\ref{calib}(c) shows the plot the microwave 
output power at the sideband peak $S_{VV}(\omega_+-\omega_m)$ 
with the mean pump photon occupation $n_d$ in the dressed mode. 
From the slope of a linear line fit, we estimate the mean 
thermal occupation $n_m$ of the mechanical resonator. 
The dressed mode frequency is tuned to $w_+/2\pi\sim 5.884$~GHz 
and axial magnetic field  $B^{\parallel}\sim18~$mT is applied.
These parameters correspond to an electromechanical coupling 
of $g_+/2\pi\sim22$~kHz, which is measured separately in CEQA 
experiment as described in the main text.
The dressed mode decay rate $\kappa/2\pi\sim 11.5 \pm
0.3$~MHz is extracted
from transmission $|S_{21}(\omega)|$.
With all these parameters, we estimate the thermal phonon 
occupation of the mechanical resonator to be $n_m\sim365$,
which corresponds to a mode temperature of $70$~mK.

\subsection{Mechanical resonator's linewidth in lower magnetic fields}

The mechanical resonator's linewidth of 13~Hz,
reported for the Device-1 in the main text is 
affected by the flux noise present in the system.
To mitigate this effect and find out the intrinsic
mechanical linewidth, we record the output mechanical
PSD of the pump while operating at a smaller magnetic
field ($B^\parallel\sim18~$mT), an operating point with the 
lower flux responsivity of the polariton mode
$(G_+/2\pi = 0.55$~GHz$/\Phi_0)$, 
and a low pump strength to avoid any backaction.

Fig.~\ref{calib}(d) shows the PSD of the lower mechanical 
sideband for a pump signal sent at $\omega_+/2\pi\sim5.864$~GHz.
By doing a Lorentzian fit on the spectrum, we determine the 
intrinsic linewidth of $\gamma_m/2\pi\sim6$~Hz.

\subsection{Estimation of the output coupling rate}

To estimate the coupling rate of the output
port with the cavity, we measure the port's
reflection $|S_{11}(\omega)|$ at 1~K temperature.
The reflection measurement is done in a 
separate cooldown where 
a 20-dB direction coupler is added to the
output line. 
The cable between the 
output port of the cavity and the directional 
coupler creates small ripples in the reflected signal
which can be seen in Fig.~\ref{calib}(e). 

We fit the data to model 
$$S_{11}(\omega)= 1- \frac{\kappa_e}{(\kappa_i+
\kappa_{in} + \kappa_e)/2+ i(\omega-\omega_c)},$$
where $\kappa_i$ is the internal cavity
decay rate, $\kappa_{in}$ is the input-port coupling rate, 
$\omega_c$ is the cavity resonance frequency and 
$\kappa_e$ is the output-port coupling rate.
From the fit, we estimate the output coupling rate to 
be $\kappa_e/2\pi\sim6.2\pm 0.1~$MHz.

\subsection{Flux-responsivity and Kerr-nonlinearity of the dressed mode}

\begin{figure*}
\includegraphics[width= 165mm]{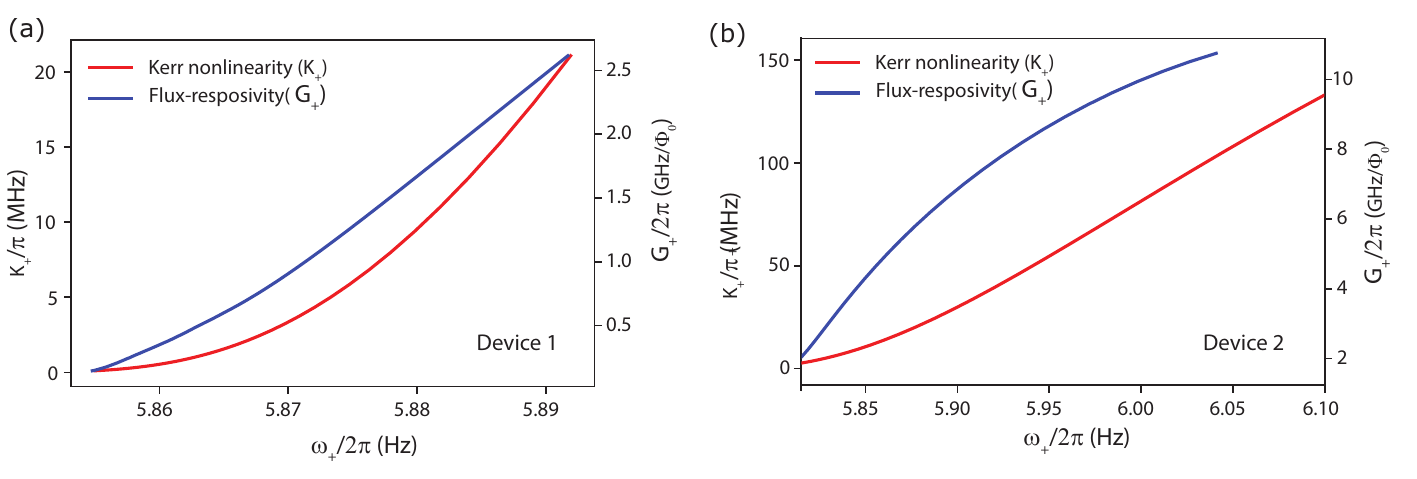}
\caption{
Plot of the flux responsivity ($G_+$) and Kerr nonlinearity 
($K_+$) of the upper polariton mode for Device-1 and Device-2 
are shown in (a) and (b), respectively. $G_+$ is extracted
from the direct cavity measurements. The anharmonicity of the 
polariton mode $|K_+|$ is numerically calculated using 
QuTip\cite{johansson_qutip:_2012} and
the experimentally determined 
device parameter given in the device parameter Table.
}
\label{FigS3}
\end{figure*}

From the cavity transmission $|S_{21}(\omega)|$ near the 
vacuum Rabi splitting, as shown in Fig.~1(d) of main text, 
we can calculate the flux responsivity of the dressed mode
for both devices.
First, we estimate the mode frequency $\omega_+$ at each 
flux bias point by fitting a Lorentzian
to $|S_{21}(\omega)|$ measurement. 
Subsequently, the extracted results can be used to 
numerically compute the first derivative of $\omega_+$ with 
respect to flux bias ($\Phi$), i.e., the flux responsivity.
In Fig.~\ref{FigS3}, we plot $G_+$ of the 
upper dressed mode with the mode frequency.

The flux responsivity can be utilized to estimate the
coupling rate at different flux bias points, employing
straight forward relation $g_+= G_+ B_\parallel x_{zpf}$.
This method is particularly useful for determining
coupling rates for mode frequencies not determined
by CEQA experiment.
From the CEQA experiment in Device-1
(Fig.~2(d) of main text), we know that 
$g_+/2\pi\approx 23.1\pm1.4~$kHz 
for dressed mode frequency 
of $5.884~$GHz when $B_\parallel$ was set to
$18~$mT.
From this known value of coupling strength, we 
estimate $g_+$
at dressed mode frequency of $5.873~$GHz to be
$13.4\pm0.8~$kHz, which is used to compute the 
black curves in Fig.~3(a) of the main text.

Next, we try to estimate the
Kerr nonlinearity of the dressed modes.
Using the QuTip package\cite{johansson_qutip:_2012},
we compute the eigen-energies of the system 
while varying the transmon qubit frequency.
The system Hamiltonian is defined using the
device parameters given in 
Table.~\ref{table-s1} and\ref{table-s2} .
Subsequently, we estimate the Kerr nonlinearity of the 
dressed mode by calculating the difference between different
energy levels, and the result is plotted in red curve
of Fig.~\ref{FigS3}.

For the numerical calculations of the Kerr nonlinearity, 
we model the transmon and cavity as a 4-level systems each. 
For the numerically calculated plot of Fig.~4(b) in the main text, the 
Hilbert space dimension is chosen to be $9$, consisting of 3-levels
of transmon and 3-levels of the cavity. 
The mean-thermal occupation is set to be 0.1 for both transmon 
and cavity, which is essential to capture the higher transitions.
These values are also used to compute the plots in Fig.~5(b) and 
(d) of the main text.

\section{}
\subsection{Modeling of cavity-enabled qubit-phonon absorption}
\subsubsection{In weak anharmonicity limit}

For the experiment discussed in Fig.~2(c) of main text,
the transmon-cavity detuning is kept such that 
the polariton mode's decay rate is larger 
than its anharmonicity.
In this limit, the upper-polariton mode
can be treated as a weak-Kerr oscillator.
Hence, only considering the upper polariton mode,
the system can be described as a weak-Kerr 
oscillator of frequency $\omega_+$ longitudinally 
coupled to a mechanical resonator of frequency $\omega_m$.
In the presence of two continuous drive signals,
a strong pump and a weak probe, 
the Hamiltonian of the system can be written as
\begin{multline}
H = \omega_+\hat{a}_+^\dag\hat{a}_+ - \frac{K_+}{2} \hat{a}_+^\dag\hat{a}_+^\dag\hat{a}_+\hat{a}_+ +
    \omega_m \hat{b}^\dag\hat{b} \\ +
    g_+~\hat{a}_+^\dag\hat{a}_+(\hat{b} + \hat{b}^\dag) 
    + \epsilon_d(\hat{a}_+e^{i\omega_dt} + \hat{a}_+^\dag e^{-i\omega_dt}) \\
    + \epsilon_p(\hat{a}_+e^{i\omega_pt} + \hat{a}_+^\dag e^{-i\omega_pt}),
\end{multline}
where $\omega_d$($\omega_p$) is the pump(probe)
drive frequency, $\epsilon_d$($\epsilon_p$) is
the strength of the signal and $K_+$ is the Kerr 
nonlinearity of porariton mode.
In the frame rotating at the control
drive frequency $\omega_d$, the Hamiltonian 
is given by
\begin{multline}
    H = - \Delta \hat{a}_+^\dag\hat{a}_+ - \frac{K_+}{2} \hat{a}_+^\dag\hat{a}_+^\dag\hat{a}_+\hat{a}_+ +
    \omega_m \hat{b}^\dag\hat{b} \\ +
    g_+~\hat{a}_+^\dag\hat{a}_+(\hat{b} + \hat{b}^\dag) 
    + \epsilon_d(\hat{a}_+ + \hat{a}_+^\dag) \\
    + \epsilon_p(\hat{a}_+e^{i\delta_pt} + \hat{a}_+^\dag e^{-i\delta_pt}),
\end{multline}
where $\Delta= \omega_d -\omega_c$, and $\delta_p =\omega_p-\omega_d$.

We ignore the quantum fluctuations and 
write down the equation of motion(EOM) 
for the mean values of the operators.
The operators $(\hat{b}+\hat{b}^{\dag})$ 
and $(\hat{b}-\hat{b}^{\dag})$ 
are essentially the normalized position and momentum,
and denoted as 
$\hat{X}$ and $\hat{P}$, respectively.
The mean value$\langle\hat{O}\rangle(t)$ of any 
operator is represented as $O$ from now on.

Then, the EOMs are given by 
\begin{subequations}
    \begin{multline}
        \stackrel{.}{a_+} = (i\Delta - \kappa/2)a_+ 
        + iK_+ a_+^\ast a_+ a_+-ig_0 a_+X \\
        -i\epsilon_d
        -i\epsilon e^{-i\delta_pt}
    \end{multline}
    \begin{multline}
        \stackrel{.}{a_+^\ast} = (i\Delta - \kappa/2)a_+^\ast
        - iK_+ a_+^\ast a_+^\ast a_+ + ig_0 a_+^\ast X \\
        +i\epsilon_d 
        + i\epsilon_p e^{i\delta_pt}
    \end{multline}
    \begin{equation}
        \stackrel{.}{ X} = -i\omega_m P
    \end{equation}
    \begin{equation}
        \stackrel{.}{P} = -i\omega_m  X
        -2ig_0 a_+^\ast a_+ - \gamma_m P.
    \end{equation}
    \label{EOM_weak_nonlinear}
\end{subequations}

To solve these equation of motions, 
we use a perturbative ansatz solution
for weak probe. It is given by
\begin{subequations}
    \begin{equation}
    a_+(t)= A_0 + A_-e^{-i\delta_pt} + A_+e^{i\delta_pt},
    \end{equation}
    \begin{equation}
    X(t)= X_0 + X_-e^{-i\delta_pt} + X_+e^{i\delta_pt},
    \end{equation}
    \begin{equation}
    P(t)= P_0 + P_-e^{-i\delta_pt} + P_+e^{i\delta_pt}.
    \end{equation}

    \label{ansatz_sol_anharmonic_OMIT}
\end{subequations}

It has a time 
independent static component along with
two other time dependent parts. 
By substituting the ansatz and comparing
the time independent components of 
Eq.~\ref{EOM_weak_nonlinear}, we arrive 
at the following steady state equations,

\begin{subequations}
    \begin{equation}
        -i\omega_m P_0 = 0
    \end{equation}
    \begin{equation}
     X_0 = -\frac{2g_0|\Bar{\alpha}|^2}{\omega_m}
    \end{equation}
    \begin{equation}
        (i\Delta - \kappa/2)A_0 + iK_+ A^\ast_0 A_0^2
        - ig_0 A_0X_0 - i\epsilon_d = 0 
    \end{equation}
    \label{anharmonic_OMIT_ss}
\end{subequations}
The Eq.~\ref{anharmonic_OMIT_ss}(a) essentially
implies that the average momentum of the 
mechanical mode is zero, while 
Eq.~\ref{anharmonic_OMIT_ss}(b) represents
the static mechanical displacement denoted by
$ X_0$.
The ``optical" mode's steady state amplitude
is represented by $A_0$, which we will henceforth
refer to as $\Bar{\alpha}$. It can be computed
by solving Eq.~\ref{anharmonic_OMIT_ss}(c).

Our goal is to determine the response $A_-$
at the probe frequency, which is experimentally
measured quantity.
In order to do this, we substitute the ansatz solutions
in Eq.~\ref{EOM_weak_nonlinear} and
compare the coefficient of $e^{-i\delta_pt}$.
Thus, we arrive at the following equations

\begin{subequations}[h]
    \begin{equation}
     B_1  X_- = -2g_0\omega_m (\Bar{\alpha}^\ast A_- + \Bar{\alpha} A^\ast_-)  \text{,} 
    \end{equation}
    \begin{equation}
       B_2 A_- = i\epsilon_p +iK_+\Bar{\alpha}^2 A^\ast_- - ig_0\Bar{\alpha} X_- \text{,}
    \end{equation}
    \begin{equation}
       B_3 A^\ast_- = - iK_+ (\Bar{\alpha}^\ast)^2 A_- + ig_0\Bar{\alpha}^\ast X_- \text{,}
    \end{equation}
    \label{anharmonic_OMIT_first_order_EOM}
\end{subequations}

where 
$$B_1= (\omega_m^2-\delta_p^2-i\gamma_m\delta_p),$$
$$B_2 = \left(\kappa/2 -i(\delta_p + \Delta) - 2iK_+|\Bar{\alpha}|^2
-2ig_0|\Bar{\alpha}|^2/\omega_m\right ) $$ 
and 
$$  B_3 = (\kappa/2 -i(\delta_p - \Delta) + 2iK_+|\Bar{\alpha}|^2
+2ig_0|\Bar{\alpha}|^2/\omega_m ).$$ 

Substituting the solution of $X_-$ from 
Eq.~\ref{anharmonic_OMIT_first_order_EOM}(a) into
Eq.~\ref{anharmonic_OMIT_first_order_EOM}(c) will
results in
\begin{equation}
    (B_3 + i|\Bar{\alpha}|^2 B_1^\prime ) A^\ast_- =
    -i(\Bar{\alpha}^\ast)^2(K_+ + B_1^\prime) A_-,
\end{equation}
where $B_1^\prime = 2g_0^2\omega_m/B_1$.
Next, we substitute the solution of $ X$
and $A^\ast_-$ into 
Eq.~\ref{anharmonic_OMIT_first_order_EOM}(b),
and arrive at the following equation

\begin{equation}
    \left(B_2 - i|\Bar{\alpha}|^2 B_1^\prime - |\Bar{\alpha}|^4\frac{(K_+ + B_1^\prime)^2}
    {B_3 + i|\Bar{\alpha}|^2 B_1^\prime}\right ) A_- = -i\epsilon_p.
\end{equation}

If the pump is applied at a red-detuned frequency,
i.e., $\omega_d = \omega_c - \omega_m$, and the 
probe is near the resonator frequency, i.e.,
$\omega_p = \omega_c + \delta$ where 
$\delta = \delta_p - \omega_m$, then the
component of the intracavity field $A_-$ at
frequency $\omega_p$ takes the following 
analytical form:
\begin{widetext}
\begin{equation}
    A_-(\delta) \approx \frac{i\epsilon_p}{ -\kappa/2 + 2iK_+|\Bar{\alpha}|^2 + 
    2i g_0|\Bar{\alpha}|^2/\omega_m + i\delta - 2|\Bar{\alpha}|^2 g_0^2/(\gamma_m - 2i\delta)
    +|\Bar{\alpha}|^4\frac{(K_+ + 2i g_0^2/(\gamma_m - 2i\delta))^2}
    {2iK_+|\Bar{\alpha}|^2 +\kappa/2 - 2i\omega_m + 2|\Bar{\alpha}|^2g_0^2/\omega_m
     (i-\omega_m/(\gamma_m- 2i\delta))}}.
\end{equation}
\end{widetext}
The experimentally measured cavity transmission
is given by $\sqrt{\kappa_e}A_-/{a_{in}}$, where
$\kappa_e$ and $a_{in}$ are the output
coupling rate and input probe strength, respectively.
The expression of $A_-$ is obtained without
the approximation of resolved sideband 
regime ($\omega_m\gg\kappa$). 
Thus, it can be used to fit the experimental data of
Fig.2(c) of the main text to obtain 
electromechanical coupling rate $g_+$.

\subsubsection{In strong anharmonicity limit}

When transmon qubit is detuned away from the
cavity frequency, its anharmonicity is not diluted
by the linear cavity and is 
large compared to the dissipation rate. 
Then, the ``transmon-like" mode can still be treated
as an effective two-level system (TLS) or qubit.
The frequency of the TLS is given by 
$\Tilde{\omega}_q= \omega_q + J^2/\Delta_q$, where
$\omega_q$, $J$ and $\Delta_q = \omega_q - \omega_c$ are 
the bare qubit frequency, transmon-cavity coupling 
rate, and detuning between transmon and cavity,
respectively. The shift in frequency arises from the
interaction with the cavity. 
Thus, the system can be described as a 
two-level system is longitudinally 
coupled to a mechanical resonator.
In the presence of a pump and a 
probe signal with frequency $\omega_d$ and $\omega_p$, 
the Hamiltonian of the system can be written as,

\begin{multline}
    H = \frac{\Tilde{\omega}_q}{2} \hat{\sigma}^z + \omega_m \hat{b}^{\dagger}\hat{b} 
    + \frac{g_0}{2}(\hat{\sigma}^z + 1)
    (\hat{b}^{\dagger} +\hat{b}) 
    + \epsilon_d (\hat{\sigma}^+ e^{-i\omega_dt}\\
    + \hat{\sigma}^-e^{i\omega_dt})
    + \epsilon_p (\hat{\sigma}^+e^{-i\omega_pt} + \hat{\sigma}^-e^{i\omega_pt})
    \label{TLS_hamiltonian}
\end{multline}
where the $\hat{\sigma}$'s are the Pauli operators 
corresponding to the TLS and 
$\hat{b}(\hat{b}^{\dagger})$ is the ladder
operator of the mechanical mode.
By shifting to a frame rotating at
pump signal's frequency $\omega_d$,
we obtain
\begin{multline}
    H= -\frac{\Delta}{2} \hat{\sigma}^z + \omega_m \hat{b}^{\dagger}\hat{b} 
    + \frac{g_0}{2}( \hat{\sigma}^z + 1)(\hat{b}^{\dagger} +\hat{b})
    \\ + \epsilon_d (\hat{\sigma}^+ + \hat{\sigma}^-) 
    + \epsilon_p (\hat{\sigma}^+e^{-i\delta_pt} + \hat{\sigma}^-e^{i\delta_pt}),
    \label{TLS_RWT_Hamiltonian}
\end{multline}

where $\Delta= \omega_d - \Tilde{\omega}_q $ and
$\delta_p = \omega_p-\omega_d$. 
The Pauli operators follow the commutation relation,
$[\hat{\sigma}^+, \hat{\sigma}^-] = \hat{\sigma}^z ,
[\hat{\sigma}^+, \hat{\sigma}^z] = -2\hat{\sigma}^+ ,$
and $[\hat{\sigma}^-, \hat{\sigma}^z] =2\hat{\sigma}^-$.

In this study, we deal with the mean 
response of the system and ignore the 
quantum fluctuation.
For simplicity, the mean value 
$\langle\hat{O}\rangle$ of an operator 
is represented as $O$.
We can construct the mean value equation 
using the Hamiltonian of
Eq.~\ref{TLS_RWT_Hamiltonian} and it is given by

\begin{subequations}
    \begin{equation}
        \stackrel{.}{ X} = -i\omega_m P,
    \end{equation}
    \begin{equation}
        \stackrel{.}{P} = -i\omega_m  X
        -ig_0(\sigma^z + 1) - \gamma_mP,
    \end{equation}
    \begin{multline}
        \stackrel{.}{\sigma}^+ = (-i\Delta-\gamma_q/2)\sigma^+ + ig_0 X\sigma^+
        -i\epsilon_d\sigma^z \\
        -i\epsilon_pe^{i\delta_pt}\sigma^z,
    \end{multline}
    \begin{multline}
        \stackrel{.}{\sigma}^- = (i\Delta-\gamma_q/2)\sigma^- - ig_0 X\sigma^-
        +i\epsilon_d\sigma^z \\
        + i\epsilon_pe^{-i\delta_pt}\sigma^z,
    \end{multline}
    \begin{multline}
        \stackrel{.}{\sigma}^z = -\gamma_q(\sigma^z + 1)
        - 2i\epsilon_d\sigma^+
        +2i\epsilon_d\sigma^- \\
        - 2i\epsilon_pe^{-i\delta_pt}\sigma^+
        + 2i\epsilon_pe^{i\delta_pt}\sigma^-.
    \end{multline}
    \label{EOM_OMIT_TLS}
\end{subequations}

Here, $\gamma_q$ and $\gamma_m$ represents
the dissipation rates of the qubit and mechanical
resonator respectively.

For a low enough strength of probe signal,
we do a perturbative expansion of the mean values
and use the Ansatz solution
\begin{equation}
    O(t)= O_0 + O_-e^{-i\delta_pt} + O_+e^{i\delta_pt},
    \label{ansatz_sol}
\end{equation}
where $O(t)$ represents the mean values of the 
operators.
The time-independent component are the steady
state amplitude, whereas the coefficient of
$e^{-i\delta_pt}$ represents the response 
at probe frequency.
By substituting the ansatz solution in
Eq.~\ref{EOM_OMIT_TLS}, and we arrive at
the steady state equations
\begin{subequations}
    \begin{equation}
        -i\omega_m P_0 = 0,
    \end{equation}
    \begin{equation}
        -i\omega_m X_0 -ig_0(\sigma^z_0 + 1) -
        \gamma_mP_0 = 0,
    \end{equation}
    \begin{equation}
        (-i\Delta-\gamma_q/2)\sigma^+_0 + ig_0 X_0 \sigma^+_0
        -i\epsilon_d\sigma^z_0 = 0,
    \end{equation}
    \begin{equation}
        (i\Delta-\gamma_q/2)\sigma^-_0 - ig_0 X_0 \sigma^-_0
        + i\epsilon_d\sigma^z_0 = 0,
    \end{equation}
    \begin{equation}
        -\gamma_q(\sigma^z_0 + 1) - 2i\epsilon_d\sigma^+_0
        +2i\epsilon_d\sigma^-_0 = 0.
    \end{equation}
\end{subequations}
From the above equation, we compute the steady 
state amplitudes 
\begin{subequations}
    \begin{equation}
        P_0 = 0,
    \end{equation}
    \begin{equation} 
        \sigma^z_0= -\frac{\Tilde{\Delta}^2+ \gamma_q^2/4}
        {{\Tilde{\Delta}}^2 +\gamma_q^2/4 + 2\epsilon_d^2},
    \end{equation}
    \begin{equation}
        \sigma^+_0 = i\epsilon_d\frac{\gamma_q/2 - i\Tilde{\Delta}}
        {{\Tilde{\Delta}}^2 +\gamma_q^2/4 + 2\epsilon_d^2},
    \end{equation}
    \begin{equation}
        \sigma^-_0 = -i\epsilon_d\frac{\gamma_q/2 + i\Tilde{\Delta}}
        {{\Tilde{\Delta}}^2 +\gamma_q^2/4 + 2\epsilon_d^2},
    \end{equation}
    \begin{align}
        X_0 & = -\frac{g_0}
        {\omega_m}(\sigma^z + 1/2)\nonumber \\
        & = -2\frac{2g_0\epsilon_d^2}{\omega_m({\Tilde{\Delta}}^2 +\gamma_q^2/4 + 2\epsilon_d^2)}.
    \end{align}
\label{TLS_OMIT_SS_sol}
\end{subequations}

Here we define $\Tilde{\Delta}=\Delta - g_0 X_0$.  

Next, we compute the first-order coefficients,
in particular $\sigma^-_-$, the quantity
that is measured experimentally.
In order to compute this, we compare the coefficients of
$e^{-i\delta_pt}$ from Eq.~\ref{EOM_OMIT_TLS}(a)
and (b), and substitute the steady-state amplitudes
from Eq.~\ref{TLS_OMIT_SS_sol}.
Thus, we arrive at the following equations

\begin{subequations}
    \begin{equation}
        -i\omega_m P_-  = -i\delta_p X_-
    \end{equation}
    \begin{equation}
        -i\delta_p P_-  = -i\omega_m X_- 
       -ig_0\sigma^z_-  -\gamma_m P_-
    \end{equation}
    \begin{multline}
        -i\delta_p\sigma^+_- = (-i\Delta-\gamma_q/2)\sigma^+_- 
    + ig_0 X_0\sigma^+_- \\
    + ig_0 X_-\sigma^+_0 - i\epsilon_d\sigma^z_-
    \end{multline}
    \begin{multline}
        -i\delta_p\sigma^-_- = (i\Delta-\gamma_q/2)
        \sigma^-_- 
    - ig_0 X_0\sigma^-_- \\
    - ig_0 X_-\sigma^-_0  + i\epsilon_d\sigma^z_-
    +i\epsilon_p\sigma^z_0
    \end{multline}
    \begin{multline}
        -i\delta_p\sigma^z_- = -\gamma_q\sigma^z_- 
    -2i\epsilon_d(\sigma^+_- - \sigma^-_-) 
    -2i\epsilon_p\sigma^+_0
    \end{multline}
    \label{TLS_OMIT_first_order}
\end{subequations}

From Eq.~\ref{TLS_OMIT_first_order}(a) and 
(b) we get
\begin{equation}
    X_- = B_4\sigma^z_-,
    \label{TLS_ss_sol_1}
\end{equation}

where $$B_4 = -\frac{ig_0}{i\omega_m + (\gamma_m
- i\delta_p )\frac{\delta_p}{\omega_m}}.$$

From Eq.~\ref{TLS_OMIT_first_order}(c) and (e) 
and using the solution of
$X_-$ form
Eq.~\ref{TLS_ss_sol_1}, we arrive at
\begin{subequations}
    \begin{equation}
        \sigma^+_-= B_5\sigma^z_- ,
    \end{equation}
    \begin{equation}
        \sigma^z_- = B_6\sigma^-_- + B_7.
    \end{equation}
\end{subequations}

Here, $$B_5 = \frac{i\epsilon_d - 
B_4\epsilon_d g_0(\gamma_q/2 - i\Tilde
{\Delta})/(\Tilde{\Delta}^2 +\gamma_q^2/4 +
2\epsilon_d^2)}{\gamma_q/2 -i
(\delta_p - \Tilde{\Delta})},$$
$$B_6= \frac{2i\epsilon_d}
{\gamma_q-i\delta_p+2i\epsilon_d B_5},$$
and $$B_7= \frac{2\epsilon_p\epsilon_d}
{(\gamma_q-i\delta_p+2i\epsilon_d B_5)}
\frac{(\gamma_q/2 - i\Tilde{\Delta})}
{(\Tilde{\Delta}^2 +\gamma_q^2/4 +
2\epsilon_d^2)}$$

Finally, from
Eq.~\ref{TLS_OMIT_first_order}(c) we compute
the analytical expression of $\sigma^-_-$.
It is given by
\begin{multline}
    B_8\sigma^-_- = \left(i\epsilon_d - \frac{g_0\epsilon_d B_4 (\gamma_q/2 +
    i\Tilde{\Delta})}{{\Tilde{\Delta}}^2 +\gamma_q^2/4 + 2\epsilon_d^2}\right)B_7
    - i\epsilon_p,
    \label{CEQA_final}
\end{multline}

where 
\begin{multline}
    B_8 = \gamma_q/2 - i(\Tilde{\Delta} +\delta_p)
- \\
\left(i\epsilon_d - \frac{g_0\epsilon_d B_4
(\gamma_q/2 + i\Tilde{\Delta})}
{{\Tilde{\Delta}}^2 +\gamma_q^2/4 + 2\epsilon_d^2}\right)B_6.
\end{multline}
Here, $\sigma^z_0$ has been approximated to be $-1$,
assuming the pump strength $\epsilon_d$ to be 
comparatively low than decay rates.
Additionally, in the experimental setup of Device-2,
it holds that $g_0$ and $\epsilon_d$ are much smaller
than $\gamma_q$. Consequently,
we can approximate $\Tilde{\Delta}\approx\Delta$.
For a red detuned pump signal, i.e.
$\Delta= \omega_d - \Tilde{\omega}_q=\omega_m$, the
expression of $\sigma_-^-$ can be computed
from Eq.~\ref{CEQA_final}.
This expression(with a normalization factor) is 
used to fit the data points in Fig.2(e) of 
the main text, resulting in the solid black curve.

\section{}
\subsection{Backaction from a weakly nonlinear-Kerr mode}

In this section, we analyze the backaction exerted
on the mechanical resonator arising due to the 
optomechanical interaction of the polariton modes.
Because of large spectral separation, we solely
focus on the upper polariton mode, leading to a simplified
two-mode analysis. 
Considering the upper polariton mode as a weak-Kerr
oscillator, in a frame rotating at the pump frequency,
the Hamiltonian of the system can be expressed as
\begin{multline}
\mathcal{H} = - \Delta \hat{a}_+^\dagger\hat{a}_+ - \frac{K_+}{2} \hat{a}_+^\dagger\hat{a}_+^\dagger\hat{a}_+\hat{a}_+ +
\omega_m \hat{b}^\dagger\hat{b} \\ +
g_+~\hat{a}_+^\dagger\hat{a}_+(\hat{b} + \hat{b}^\dagger) + 
\epsilon~(\hat{a}_+ + \hat{a}_+^\dagger),
\label{rotated hamiltonian}
\end{multline}
where $K_+$ is the Kerr-nonlinearity of
the upper polariton mode,
$\epsilon$ is the drive strength of the pump,
$\Delta=\omega_d-\omega_+$ is the drive detuning, 
and the rest of the symbols have their usual meaning.

The quantum Langevin equations of the system are given by
\begin{subequations}
\begin{equation}
    \dot {\hat{a}}_+ = - {i}\left[ \hat{a}_+, \mathcal{H} \right] - \frac{\kappa}{2} \hat{a}_+ + \sqrt{\kappa_{ex}}~\hat{a}_{in} + \sqrt{\kappa_{0}}~\hat{f}_{in},
\end{equation}
\begin{equation}
    \dot {\hat{b}} = - {i}\left[ \hat{b}, \mathcal{H} \right] - \frac{\gamma}{2} \hat{b} + \sqrt{\gamma_m}~\hat{b}_{in},
\end{equation}
\label{EOM_1}
\end{subequations}
where $\hat{a}_{in}$, $\hat{f}_{in}$ are noise operators of the 
polariton mode and  $\hat{b}_{in}$ is noise operator 
of the mechanical mode,
$\kappa(\gamma_m)$ is the decay rate of the polariton (mechanical) 
mode, and 
$\kappa_{ex} (\kappa_{0})$ are the total external(internal)
decay rates of the polariton mode, respectively.
Using the Hamiltonian in Eq.~\ref{rotated hamiltonian} and Eq.~\ref{EOM_1},
we obtain the equations of motion (EOM) of both modes:

\begin{subequations}
\begin{multline}
    \dot {\hat{a}}_+ = ({i}\Delta - \frac{\kappa}{2})\hat{a}_+ + {i}K_+\hat{a}_+^\dagger \hat{a}_+ \hat{a}_+ - {i} g_0 \hat{a}_+ (\hat{b} + \hat{b}^\dagger) \\
    + \epsilon + \sqrt{\kappa_{ex}}~\hat{a}_{in} + \sqrt{\kappa_{0}}~\hat{f}_{in},
\end{multline}
\begin{equation}
    \dot {\hat{b}} = (-{i}\omega_m - \frac{\gamma_m}{2})\hat{b} - {i}g_0 \hat{a}_+^\dagger \hat{a}_+ + \sqrt{\gamma_m}~\hat{b}_{in}.
\end{equation}
\label{full_equation_of_motion}%
\end{subequations}

By defining the mean field occupation of the
polariton mode and the mechanical mode as 
$\alpha$ and $\beta$, respectively,
we arrive at the following semi-classical 
equations of motion:

\begin{subequations}
\begin{equation}
    \dot{\alpha} = ({i}\Delta - \frac{\kappa}{2})\alpha + {i} K_+ \left|\alpha \right|^2 \alpha - {i} g_0 \alpha (\beta + \beta^\ast) + \epsilon,
\end{equation}
\begin{equation}
    \dot{\beta} = (-{i}\omega_m - \frac{\gamma_m}{2}) \beta - {i}g_0 \left|\alpha \right|^2.
\end{equation}
\label{steady_state_equations}
\end{subequations}
In the steady state of the system, these equations become
$\dot{\alpha} = 0$, $\dot{\beta} = 0$. Solving these, we find 
the steady state field amplitude $\Bar{\alpha}$ 
and $\Bar{\beta}$.

Next, we assume an ansatz solution of 
Eq.~\ref{full_equation_of_motion}, where we write down the
fields as a combination of steady state 
amplitude and a fluctuation term, \textit{i.e.}
$\hat{a}_+ = \Bar{\alpha} + \delta\hat{a}_+$, and 
$\hat{b} = \Bar{\beta} + \delta\hat{b}$. 
Substituting these ansatz to Eq.~\ref{full_equation_of_motion}, 
and ignoring the higher order fluctuation terms,
we get the EOMs as,

\begin{subequations}
\begin{multline}
        \dot{\delta\hat{a}}_+ = \left[{i}\Tilde{\Delta} - \frac{\kappa}{2}\right]\delta\hat{a}_+ + {i}\eta \delta\hat{a}_+^\dagger  - {i} G (\delta\hat{b} + \delta\hat{b}^\dagger) + \sqrt{\kappa_{ex}}~\hat{a}_{in} \\
        + \sqrt{\kappa_{0}}~\hat{f}_{in},
\end{multline}
\begin{multline}
\dot{\delta\hat{a}}_+^\dagger = \left[-{i}\Tilde{\Delta} - \frac{\kappa}{2}\right]\delta\hat{a}_+^\dagger - {i}\eta^\ast \delta\hat{a}_+  + {i} G^\ast (\delta\hat{b} + \delta\hat{b}^\dagger) + \\
        \sqrt{\kappa_{ex}}~\hat{a}_{in}^\dagger + \sqrt{\kappa_{0}}~\hat{f}_{in}^\dagger,
\end{multline}
\begin{multline}
        \dot{\delta\hat{b}} = \left[-{i}\omega_m - \frac{\gamma_m}{2}\right]\delta\hat{b} - {i}G^\ast \delta\hat{a}_+
        - {i}G \delta\hat{a}_+^\dagger + \sqrt{\gamma_m}~\hat{b}_{in},
\end{multline}
\begin{multline}
        \dot{\delta\hat{b}}^\dagger = \left[{i}\omega_m - \frac{\gamma_m}{2}\right]\delta\hat{b}^\dagger
        + {i}G \delta\hat{a}_+^\dagger + {i}G^\ast \delta\hat{a}_+ \\
        + \sqrt{\gamma_m}~\hat{b}_{in}^\dagger,
\end{multline}
\label{EOM_2}
\end{subequations}
where $\Tilde{\Delta} = \Delta +  2K_+ \left|\Bar{\alpha} \right|^2 - 
g_0 (\Bar{\beta} + \Bar{\beta}^\ast)$, $G = g_0 \Bar{\alpha}$ and $\eta = K_+ \Bar{\alpha}^2$.

Next, we perform Fourier transform of the above 
equations by defining the 
transformation as $x[\omega] = \mathcal{F}\left[x(t)\right] = \int_{-\infty}^{+\infty} x(t) \mathrm{e}^{{i}\omega t} \mathrm{d}t$.
Using the identities $(x^\dagger)[\omega] = (x[-\omega])^\dagger$, and
$\mathcal{F}[\dot{x}(t)] = -{i}\omega \mathcal{F}[x(t)]$, the new set
of equations of motion in frequency domain become,

\begin{widetext}
\begin{equation}
    \begin{bmatrix}
        \chi_c^{-1} & - {i}\eta \\ 
        {i}\eta^\ast & \Tilde{\chi}_c^{-1}
    \end{bmatrix}
    \begin{bmatrix}
        \delta\hat{a}_+[\omega] \\ 
        (\delta\hat{a}_+^\dagger)[\omega]
    \end{bmatrix} = - {i}
    \begin{bmatrix}
        G & G \\ 
        -G^\ast & -G^\ast
    \end{bmatrix}
    \begin{bmatrix}
        \delta\hat{b}[\omega] \\ 
        (\delta\hat{b}^\dagger)[\omega]
    \end{bmatrix} + 
    \begin{bmatrix}
        \sqrt{\kappa_{ex}}~\hat{a}_{in}[\omega] + \sqrt{\kappa_{0}}~\hat{f}_{in}[\omega] \\ 
        \sqrt{\kappa_{ex}}~(\hat{a}_{in}^\dagger)[\omega] + \sqrt{\kappa_{0}}~(\hat{f}_{in}^\dagger)[\omega]
    \end{bmatrix}
    \label{cavity_matrix_eq}
\end{equation}

\begin{equation}
    \begin{bmatrix}
        \chi_m^{-1} & 0 \\ 
        0 & \Tilde{\chi}_m^{-1}
    \end{bmatrix}
    \begin{bmatrix}
        \delta\hat{b}[\omega] \\ 
        (\delta\hat{b}^\dagger)[\omega]
    \end{bmatrix} = - {i}
    \begin{bmatrix}
        G^\ast & G \\ 
        -G^\ast & -G
    \end{bmatrix}
    \begin{bmatrix}
        \delta\hat{a}_+[\omega] \\ 
        (\delta\hat{a}_+^\dagger)[\omega]
    \end{bmatrix} + 
    \begin{bmatrix}
        \sqrt{\gamma_m}~\hat{b}_{in}[\omega] \\ 
        \sqrt{\gamma_m}~(\hat{b}_{in}^\dagger)[\omega]
    \end{bmatrix},
    \label{mech_matrix_eq}
\end{equation}
\end{widetext}
where $\chi_c^{-1}[\omega] = (-{i}(\omega + \Tilde{\Delta}) + \kappa/2)$, 
$\Tilde{\chi}_c^{-1}[\omega] = (-{i}(\omega - \Tilde{\Delta}) + \kappa/2)$ are dressed mode's susceptibilities and
$\chi_m^{-1}[\omega] = (-{i}(\omega - \omega_m) + \gamma_m/2)$, 
$\Tilde{\chi}_m^{-1}[\omega] = (-{i}(\omega + \omega_m) + \gamma_m/2)$ 
is the mechanical susceptibilities. 

To find the effective dynamics of the mechanical resonator, we
first solve Eq.~\ref{cavity_matrix_eq}, and substitute the solution
of $(\delta\hat{a}_+)[\omega]$ and
$(\delta\hat{a}_+^\dagger)[\omega]$ in 
Eq.~\ref{mech_matrix_eq}. This leads to the simplified equations
of the mechanical mode as, 

\begin{widetext}
\begin{equation}
    \begin{bmatrix}
        -{i}(\omega - \omega_m) + \frac{\gamma_m}{2} + \Sigma_c[\omega] & 
        \Sigma_c[\omega] \\ 
        -\Sigma_c[\omega] & -{i}(\omega + \omega_m) + \frac{\gamma_m}{2} -\Sigma_c[\omega]
    \end{bmatrix}
    \begin{bmatrix}
        \delta\hat{b}[\omega] \\ 
        (\delta\hat{b}^\dagger)[\omega]
    \end{bmatrix} =
    \sqrt{\gamma_m}
    \begin{bmatrix}
        \hat{B}_{in}[\omega] \\ 
        (\hat{B}_{in}^\dagger)[\omega]
    \end{bmatrix}.
    \label{backaction}
\end{equation}
\end{widetext}

\begin{figure*}
\centering
\includegraphics[width= 165mm]{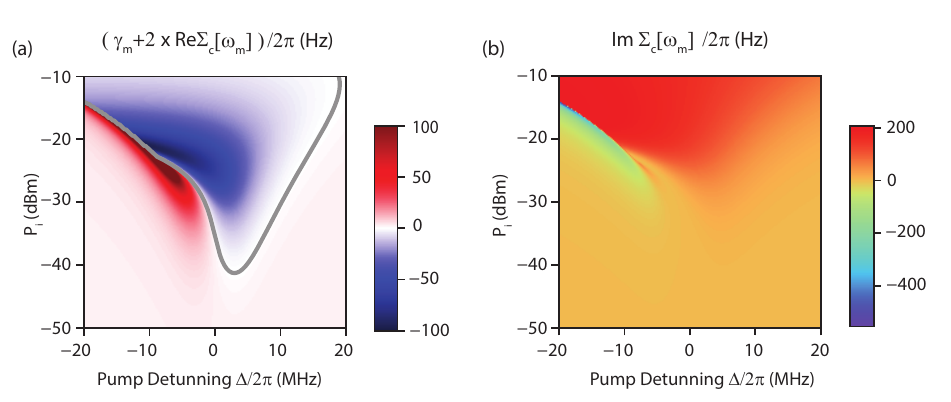}
\caption{\textbf{Backaction from weak Kerr oscillator model:}
Colorplot in (a) and (b) shows the effective mechanical
linewidth $\Gamma_m = \gamma_m + 
2\times$Re({$\Sigma_c[\omega_m]$})
and optomechanical frequency shift
$\delta\omega_m =$ Im({$\Sigma_c[\omega_m]$}), 
respectively.
To maintain the visibility of small variations, the
color scale in (a) is truncated at $\pm 100~$Hz.
The gray curve in (a) denotes the contour
of $\Gamma_m=0$.
These quantities are
computed from a model where the EM mode is considered
as an anharmonic oscillator, and it is longitudinally
coupled to the mechanical resonator. The parameters
used to compute these quantities are taken from device-2,
and they are
given by $\gamma_m/2\pi\sim 6~$Hz, $\omega_m/2\pi\sim
3.97$~kHz, $K_+/2\pi\sim 8.55~$MHz, $\kappa/2\pi\sim 9~$
MHz and $g_+/2\pi\sim 45~$kHz.
}
\label{FigS4}
\end{figure*}

The quantity defined as
$\Sigma_c[\omega] =  2 {i} |G|^2 [ \Tilde{\Delta} - |\eta|][ 1/\chi_c \Tilde{\chi}_c - |\eta|^2 ]^{-1}$ 
represents the modification in the mechanical resonator's 
dynamics due to nonlinear Kerr mode. 
The frequency shift and the effective 
optomechanical damping rate of 
the mechanical resonator is given by,
$\delta\omega_m =$ Im({$\Sigma_c[\omega_m]$})
and $\Gamma_m = \gamma_m + 2\times$Re({$\Sigma_c[\omega_m]$}),
respectively. 
These two quantities are plotted for 
Device-2 parameters in Fig.~\ref{FigS4} as 
a function of pump strength and detuning. 
It illustrates how the backaction effect evolves
as parameters change. 

In Fig.~3(a) of the main text, we plot
$\Gamma_m$ and $\delta\gamma_m$ as solid black
lines using Device-1's parameters.
Similarly, by using the parameters of Device-2,
the boundary of mechanical instability is
derived from the threshold 
Re$(\Sigma_c[\omega]) = -\gamma_m/2$
and it is plotted as the solid black line
in Fig.~3(c) of the main text.
The same is shown in Fig.~\ref{FigS4} as the
gray curve.
By replacing $K_+=0$ in the expression of 
$\Sigma_c[\omega_m]$, we can calculate 
the boundary of mechanical instability for a
linear resonator coupled to the mechanical resonator.
It is shown as the dashed curve in Fig.~3(c) of
the main text.

\section{}
\subsection*{Linear stability test using semi-classical
analysis of three-mode system}

We start with the Hamiltonian of the three-mode
system consisting of a linear cavity, transmon qubit,
and a mechanical resonator.
In the presence of a drive signal, it can be written 
as
\begin{multline}
H = \omega_c a^\dagger a + \omega_q c^\dagger c - \frac{\alpha_T}{2} c^\dagger c^\dagger c c + J 
(a c^\dagger + a^\dagger c) + \omega_m b^\dagger b \\
+ g_0 c^\dagger c (b+b^\dagger) + \epsilon (a 
e^{i\omega_d t} +a^\dagger e^{-i\omega_dt}),
\end{multline}
where $J$ is
the coupling strength between cavity and transmon, 
$g_0$ is the electromechanical coupling between the transmon and the 
mechanical resonator, $\alpha_T$ is the transmon anharmonicity,
$a(a^\dagger), c(c^\dagger) $ and 
$b(b^\dagger)$ are the annihilation(creation) 
operators of cavity, transmon, and mechanical
modes with resonant frequencies of $\omega_c$, $\omega_q$ 
and $\omega_m $, respectively. 
A pump signal is continuously applied to the cavity
with strength $\epsilon$ and frequency $\omega_d$.
In the rotating frame of pump frequency, the
Hamiltonian becomes
\begin{multline}
    H = -\Delta_1 a^\dagger a - \Delta_2 c^\dagger c - \frac{\alpha_c}{2} c^\dagger c^\dagger c c + J 
    (a c^\dagger + a^\dagger c)\\ + \omega_m b^\dagger b 
    + g_0 c^\dagger c (b+b^\dagger) + \epsilon (a +a^\dagger),
\end{multline}
where $\Delta_1 = (\omega_d-\omega_c)$ 
and $\Delta_2 = (\omega_d-\omega_q)$.

Writing the Heisenberg-Langevin equations and using semi-classical 
approximation, we get the steady state equation of motions as
\begin{subequations}
\begin{equation}
    \dot{\alpha}^\prime = - (\kappa_b/2 - i\Delta_1)\alpha^\prime-
    iJ\zeta-i\epsilon ,
\end{equation}
\begin{equation}
    \dot\zeta = -[\gamma/2-i\Delta_2-
    2i\alpha_T|\zeta|^2+ i g_0\zeta
    (\beta+\beta^*)] 
    -i J\alpha^\prime,
\end{equation}
\begin{equation}
    \dot{\beta} = -(-\gamma_m/2 +
    i\omega_m)\beta -
    i g_0 \vert\zeta\vert^2,
\end{equation}
\end{subequations}
where $\langle a\rangle = \alpha^\prime$,
$\langle c\rangle = \zeta$ and 
$\langle b\rangle = \beta $ are the mean
values.

Subsequently, representing the steady-state
amplitudes in complex form as $\alpha^\prime = x+i y,
\zeta = p + i q,
$ and $\beta = u + i v$, we get the following
sets of equations
\begin{subequations}
    \begin{equation}
        \dot x = \dot f_1 = -\frac{\kappa_b}{2} x - \Delta_1 y + J q
    \end{equation}
    \begin{equation}
        \dot y = \dot f_2 = +\Delta_1 x - \frac{\kappa_b}{2} y - J p -\epsilon
    \end{equation}
    \begin{equation}
        \dot p = \dot f_3 = -\frac{\gamma}{2} p + \left(-\Delta_2 - 2\alpha_T(p^2+q^2)+2 g_0 u \right) q +J y
    \end{equation}
    \begin{equation}
        \dot q = \dot f_4 = -\frac{\gamma}{2} q - \left(-\Delta_2 - 2\alpha_T(p^2+q^2) + 2 g_0 u \right) p - J x 
    \end{equation}
    \begin{equation}
        \dot u = \dot f_5 = -\frac{\gamma_m}{2} u + \omega_m v
    \end{equation}
    \begin{equation}
        \dot v = \dot f_6 = -\omega_m u - \frac{\gamma_m}{2} v -g_0 (p^2+q^2)
    \end{equation}
\end{subequations}

For the steady-state solution or the fix point of
the system, we set the first derivatives to zero
i.e., $\dot x = \dot y = \dot p = \dot q = \dot
u = \dot v = 0$. 
It leaves us with
\begin{equation}
    \frac{\gamma_m^2}{4\omega_m} u + \omega_m u - g_0 \left( \frac{J\kappa_b\epsilon}{2 A B} p(u) + \frac
    {J\Delta_1\epsilon}{A B} q(u) \right) = 0,
    \label{eq:steady_state_equation}
\end{equation}
where $$A = \frac{\kappa_b^2}{4}+\Delta_1^2$$ 
$$B = \left( \frac{\gamma}{2} + \frac{J^2\kappa_b}{2 A} \right),$$
$$C = -\Delta_2-2\alpha_T(p^2+q^2)+2 g_0 u + \frac{J^2\Delta_1}{A}.$$
Here,
$$p(u) = -\frac{\left(\frac{J\Delta_1 C}{A}+\frac{J\kappa_b B}{2 A}\right) \epsilon}{B^2+C^2},$$ 
and 
$$q(u) =  \frac{\left(\frac{-J\Delta_1 B}{A}+\frac{J\kappa_b C}{2 A}\right) \epsilon}{B^2+C^2}.$$

By finding the roots of Eq.\ref{eq:steady_state_equation}, the 
fixed points $x, y, p, q, u$ and $v$ can be obtained.
Considering the steady-state values as ($\Bar{x},\Bar{y},\Bar{p},\Bar{q},
\Bar{u},\Bar{v}$), the nature of these points can be understood by
perturbing these points and finding the time evolution of the 
perturbation.
Defining the perturbation as $z_i = k_i - \Bar{k_i}$, where
$k_i$ for $i=1,2,3,4,5,6 $ corresponds to $(x,y,p,q,u,v)$, respectively.
Subsequently, the time evolution of perturbation
$z_i$ is obtained as,
\begin{equation}
\dot z_i = \dot k_i \approx f_i\vert_{\Bar{k}} + \sum_i (k_i - \Bar{k}_i) \frac{\partial f_i}{\partial k_i}\vert_{\Bar{k}_i}.
\end{equation}
This gives

\begin{equation}
\frac{d}{dt}\left[\begin{array}{l}
z_1 \\
z_2 \\
z_3 \\
z_4\\
z_5\\
z_6
\end{array}\right]=\left[\begin{array}{llllll}
\frac{\partial f_1}{\partial x} & \frac{\partial f_1}{\partial y} & \frac{\partial f_1}{\partial p} & \frac{\partial f_1}{\partial q}& \frac{\partial f_1}{\partial u}& \frac{\partial f_1}{\partial v} \\
\frac{\partial f_2}{\partial x} & \frac{\partial f_2}{\partial y} & \frac{\partial f_2}{\partial p} & \frac{\partial f_2}{\partial q} & \frac{\partial f_2}{\partial u}& \frac{\partial f_2}{\partial v}\\
\frac{\partial f_3}{\partial x} & \frac{\partial f_3}{\partial y} & \frac{\partial f_3}{\partial p} & \frac{\partial f_3}{\partial q}& \frac{\partial f_3}{\partial u}& \frac{\partial f_3}{\partial v} \\
\frac{\partial f_4}{\partial x} & \frac{\partial f_4}{\partial y} & \frac{\partial f_4}{\partial p} & \frac{\partial f_4}{\partial q}& \frac{\partial f_4}{\partial u}& \frac{\partial f_4}{\partial v}\\
\frac{\partial f_5}{\partial x} & \frac{\partial f_5}{\partial y} & \frac{\partial f_5}{\partial p} & \frac{\partial f_5}{\partial q}& \frac{\partial f_5}{\partial u}& \frac{\partial f_5}{\partial v}\\
\frac{\partial f_6}{\partial x} & \frac{\partial f_6}{\partial y} & \frac{\partial f_6}{\partial p} & \frac{\partial f_6}{\partial q}& \frac{\partial f_6}{\partial u}& \frac{\partial f_6}{\partial v}\\
\end{array}\right]\left[\begin{array}{c}
z_1 \\
z_2 \\
z_3 \\
z_4\\
z_5 \\
z_6
\end{array}\right] .
\end{equation}

Upon substituting the values of $f_i$ 's and evaluating the derivative at the 
steady state points, we get
\begin{equation}
\frac{d}{d t}\left[\begin{array}{l}
z_1 \\
z_2 \\
z_3 \\
z_4\\
z_5\\
z_6
\end{array}\right]=S\left[\begin{array}{l}
z_1 \\
z_2 \\
z_3 \\
z_4 \\
z_5\\
z_6
\end{array}\right] .
\label{time_evolution}
\end{equation}

The $S$-matrix governs the evolution of the perturbation, and it is 
given by
\begin{widetext}
\begin{equation}
S = \left[\begin{array}{cccccc}
-\frac{\kappa_b}{2} & -\Delta_1 & 0 & J & 0 & 0 \\
\Delta_1 & -\frac{\kappa_b}{2} & -J & 0 & 0 & 0 \\
0 & J & -\frac{\gamma}{2}-4 \alpha_T \Bar{p} \Bar{q} & -\Delta_2-2 \alpha_T \Bar{p}^2-6 \alpha_T \Bar{q}^2+2 g_0 \Bar{u} & 2 g_0 \Bar{q} & 0 \\
-J & 0 & \Delta_2+6 \alpha_T \Bar{p}^2+2 \alpha_T \Bar{q}^2-2 g_0 \Bar{u} & -\frac{\gamma}{2}+4 \alpha_T \Bar{p} \Bar{q} & -2 g_0 \Bar{p} & 0 \\
0 & 0 & 0 & 0 & -\frac{\Gamma}{2} & \omega_m \\
0 & 0 & -2 g_0 \Bar{p} & -2 g_0 \Bar{q} & -\omega_m & -\frac{\Gamma}{2}
\end{array}\right] .
\end{equation}
\end{widetext}

The Eq.~\ref{time_evolution} has the solution of the form 
$z(t)=\sum_{i} b_i w_i e^{\lambda_i t}$, where $b_i$'s are
the constant of integration, $\lambda_i$'s are 
the eigenvalues of the matrix $S$ and $w_i$'s are the corresponding eigenvectors.
Thus, any eigenvalue of the $S$ with a positive real part 
will cause the solution for $z(t)$ to grow exponentially, 
resulting in instability. Thus, this becomes a criterion for identifying the unstable points.

Fig.~6(c) of the main text shows the result 
of such a calculation for Device-2. 
The parameters used for the calculation
are mentioned below.
A bare cavity decay rate of $\kappa_b\sim 8~$MHz and
transmon dissipation rate of $\gamma\sim 12~$MHz is used for
the calculation. 
The electromechanical coupling
between the transmon and mechanical resonator
is set to $300~$kHz, which was estimated
from the upper-polariton mode's flux responsivity,
given in Fig.~\ref{FigS3}.
It is evident that a semi-classical description
of the system is not sufficient to understand
the experimental observation of mechanical
parametric instability. 

\section{}
\subsection*{Modelling  of the instability region using polariton basis}

We find out in the previous section that a classical description 
of the system fails to describe the experimental observation. 
Therefore, a quantum mechanical description of electrical modes 
is necessary to explain the observations at low
to moderate pump powers.
This is done by treating each transition
into their two-level subspace.
Alternatively one can treat the electrical modes
as a multi-level atom. However, such an analysis 
quickly becomes intractable.
We justify the validity of two-level model using the fact that 
a pump near a certain transition frequency only drive
that particular transition occurs due to the large 
spectral separation compared to their decay rates.
In addition, these transitions are flux tunable
as shown in the Fig~.4(a) of the main text, resulting
in coupling with the mechanical resonator.
Thus, the full system can be treated as a multiple two-level 
systems (TLS) independently coupled to the mechanical resonator 
with a certain coupling strength.
We separately compute the region of mechanical instability 
for each TLS and superpose them together to compute
the full instability phase space diagram.

\begin{figure}[h]
    \centering
    \includegraphics[width= 70 mm]{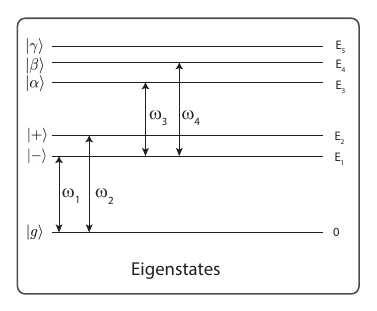}
    \caption{\textbf{Eigenstates:}
    Energy eigenstates of the cavity-transmon 
    system.  The energies are indicated on 
    the right, while the eigenstates are 
    labeled on the left.
    $\omega_i$'s represent the transition energies,
    and the arrows point to the appropriate 
    eigenstates for the specific transition.
    }
    \label{FigS5}
\end{figure}

We begin the theoretical analysis by computing the 
frequency $\omega_i$ and electromechanical
coupling strength $g_i$ for each TLS.
The transition frequencies $\omega_i$'s are obtained
from the difference
of eigenenergies of the transmon-cavity system.
The Hamiltonian of the transmon-cavity system (ignoring
the mechanical resonator) is given by
$H = \omega_c a^\dagger a + \omega_q 
c^\dagger c - \frac{\alpha_T}{2} c^\dagger
c^\dagger c c + J (a c^\dagger + a^\dagger 
c)$, 
where $\hat{a}$ and $\hat{c}$ are the ladder operators
of cavity and transmon, respectively.
We write the Hamiltonian up to two-excitation subspace 
and subsequently diagonalize it to find the energy 
eigenvalues. It is given by
\begin{widetext}
\begin{equation}
H = \left[\begin{array}{cccccc}
0 & 0 & 0 & 0 & 0 & 0 \\
0 & \omega_q & J & 0 & 0 & 0 \\
0 & J & \omega_c & 0 & 0 & 0 \\
0 & 0 & 0 & 2\left(\omega_q-\alpha_c\right) & \sqrt{2} J & 0 \\
0 & 0 & 0 & \sqrt{2} J & \omega_c+\omega_q & \sqrt{2 J} \\
0 & 0 & 0 & 0 & \sqrt{2} J & 2 \omega_c
\end{array}\right] 
\xrightarrow[]{\text{Diagonalization}}
\left[\begin{array}{cccccc}
0 & 0 & 0 & 0 & 0 & 0 \\
0 & E_1 & 0 & 0 & 0 & 0 \\
0 & 0 & E_2 & 0 & 0 & 0 \\
0 & 0 & 0 &E_3 & 0 & 0 \\
0 & 0 & 0 & 0 & E_4 & 0 \\
0 & 0 & 0 & 0 & 0 & E_5
\end{array}\right],
\end{equation}
\end{widetext}
where $E_i$'s are a function of $\omega_q, 
\omega_c, \alpha_T$ and $J$.
Using these energy eigenvalues, we obtain the 
transition frequencies $\omega_i$'s by 
calculating the difference between the relevant
energy eigenvalues, as shown by the arrows in
Fig.~\ref{FigS5}.
In terms of notation used in the main text,
the frequencies
$\omega_-,\omega_+,\omega_{-\alpha}$ and $\omega_{-\beta}$ 
correspond to
$\omega_1,\omega_2,\omega_3$ and $\omega_4$,
respectively.

Next, we determine the electromechanical coupling strengths
for each transition.
In the presence of magnetic field the transmon frequency 
becomes a function of mechanical displacement $x$, 
i.e. $\omega_q(x) \approx \omega_q^{bare} + G_q x $.
Consequently, the transition frequencies become 
a function of mechanical
displacement as well, since they rely on $\omega_q$.
Therefore, by doing Taylor's expansion of $\omega_i$'s 
up to first order in $x$, we get 
\begin{equation}
\omega_i \approx \omega_i|_{x=0} + G_i^\prime x.
\end{equation}

Here, $G_i^\prime= \left.\frac{\partial \omega_i}
{\partial x}\right|_{x=0}$ is the frequency shift
per unit displacement. 
Thus, the electromechanical coupling strength
for i'th transition is then given by
\begin{equation}
    g_i = \frac{\partial\omega_i}{\partial x}x_{zpf} =
    \frac{\partial\omega_i}{\partial\Phi}
    \frac{\partial\Phi}{\partial x}x_{zpf}
    = G_i B^\parallel l x_{zpf} \text{,}
    \label{g_i}
\end{equation}
where
$G_i=d\omega_i/d\Phi$ is the flux responsivity.
Since $G_i$'s are a function of $\omega_q$, 
we can estimate the remainder
by computing the value of any one of the $G_i$.
Here, $G_2$ is essentially the flux
responsivity of the upper polariton mode, 
and it is measured experimentally, as shown in 
Fig.~\ref{FigS3}. From this known value of
$G_2$ we calculate the remaining $G_i$'s
and hence the coupling strength $g_i$. 

The Hamiltonian of any specific two-level
system takes the form (in the interaction picture)

\begin{equation}
    H= -\Delta_i \frac{\sigma^z_i}{2} + \omega_m b^\dagger b +\frac{ g_i}{2} (\sigma^z_i + 1)(b + b^\dagger) + \epsilon_i (\sigma^+_i + \sigma^-_i),
\end{equation}

where $\Delta_i = (\omega_d-\omega_i)$,
$\omega_d$ is the drive frequency 
and $\omega_i$ is the frequency 
of the i'th transitions,  
$g_i$ is the single photon electromechanical 
coupling , and $\epsilon_i$ is the drive
amplitude.

In order to find the phase diagram of unstable 
response of the mechanical resonator, we follow the same 
approach as described in the previous section.
It starts with writing the Heisenberg-Langevin
equation for of $\sigma$'s and b, followed by 
deriving the steady state equation of motion
of the mean values of the operators:

\begin{subequations}
    \begin{equation}
        \dot\sigma^z_i = -i (2\epsilon\sigma^+ - 2\epsilon\sigma^-) - \gamma_i(\sigma^z_i+1)
    \end{equation}
    \begin{equation}
        \dot\sigma^-_i = -i(-\Delta_i\sigma^-_i + G_i x \sigma^-_i - \epsilon\sigma^z_i) - (\frac{\gamma_i}{2}+\gamma^\phi_i)\sigma^-_i
    \end{equation}
    \begin{equation}
        \dot\sigma^+_i = i (-\Delta_i\sigma^+_i + G_i x \sigma^+_i - \epsilon\sigma^z_i) - (\frac{\gamma_i}{2}+\gamma^\phi_i)\sigma^+_i
    \end{equation}
    \begin{equation}
        \dot b = -i\frac{g_i}{2}(\sigma^z_i+1)+ \omega_m b -\frac{\gamma_m}{2}b
    \end{equation}
\end{subequations}

Here we use the notations $\sigma$ and $b$ in place of
of $\langle\sigma\rangle$ and $\langle b\rangle$ to represent
the mean values. $\gamma_i$ and $\gamma_\phi$
are the energy dissipation rate and the dephasing rate
of the i'th transition.
We write the mean values in complex form as
$\sigma^z = s,~ \sigma^+ = p^\prime + i q^\prime, \text{ and }
b = u + i v$.

It is followed by 
the calculation of fixed points, which is carried out
by setting the first time derivative of the mean values
to zero. It leaves us with

\begin{equation}
    \left(\frac{\gamma_i}{2} + \gamma^\phi_i \right) q^\prime(s) = -\Delta_i p^\prime(s) + 2 g_i u(s) p^\prime(s) - \epsilon s,
\end{equation}
where 
$$q^\prime(s) = \frac{\gamma_i}{4\epsilon}(s+1),$$
$$u(s) = -\frac{\frac{g}{2}(s+1)}{\frac{\gamma_m^2}{4\omega_m}+\omega_m},$$
$$v(s) = -\frac{\frac{g\gamma_m}{4\omega_m}(s+1)}{\frac{\gamma_m^2}{4\omega_m}+\omega_m},$$
and
$$p^\prime(s) = -q^\prime(s) \frac{-\Delta_i+2 g u(s) }{\gamma_i/2+\gamma^\phi_i}.$$

Subsequently, we determine the nature of these
fixed points by finding the time evolution 
of a small perturbation. Following the same 
approach given in the previous section,
we compute the evolution matrix of the perturbation.
It is given by,

\begin{equation}
S = \left[\begin{array}{ccccc}
-\gamma_i & 0 & 4\epsilon_i & 0 & 0  \\
0 & -(\gamma_i/2+\gamma^\phi_i) & \Delta_i - 2 g u & -2 g q^\prime & 0  \\
-\epsilon & \Delta+2 g u & -(\gamma_i/2+\gamma^{\phi}_i) & 2 g p^\prime & 0  \\
0 & 0 & 0 & \frac{-\gamma_m}{2} & \omega_m\\
\frac{-g}{2} & 0 & 0 & -\omega_m & \frac{-\gamma_m}{2} \\

\end{array}\right] 
\end{equation}
If the real part of the eigenvalue
becomes positive for a certain value
of $\Delta$ and $\epsilon$, we denote that point 
in the phase space as unstable.
Hence, we compute the mechanical instability 
phase diagram of the four relevant 
two-level systems with frequencies
$\omega_1, \omega_2, \omega_3 $ and $ \omega_4$,
respectively.
Subsequently, we plot all four regions together,
resulting in the green shaded area in Fig.~6(b) 
of the main text.

It is evident from the transmission spectrum of Fig.~4(b)
in the main text that the linewidths associated 
with each transition are not equal.
The higher-level transitions have larger linewidth
compared to the lower-level transitions.
The energy decay rates and dephasing rates used
to compute Fig.~6(b) of the main text are given by
$\gamma_1\sim10~$MHz, $\gamma_1^\phi\sim4~$MHz,
$\gamma_2\sim10~$MHz, $\gamma_2^\phi\sim4~$MHz,
$\gamma_3\sim18~$MHz, $\gamma_3^\phi\sim8~$MHz,
$\gamma_4\sim14~$MHz, and $\gamma_4^\phi\sim9~$MHz.
The onset of instability for each transition 
depends on these decay rates, as observed in 
the Fig~6(b) of the main text.

In addition, the onset of instability
depends on the thermal occupation of the eigenstates.
The ground and excited state occupation of a certain TLS 
determines the probability of transition when subjected
to a drive. 
Since $\ket{+}$ and $\ket{-}$ have much smaller
thermal occupations than $\ket{g}$, the higher transitions
with frequencies of $\omega_3$ and $\omega_4$ 
are less likely to occur than lower transitions
with frequencies of $\omega_1$ and $\omega_2$.
While computing the instability boundary, we 
consider 82\% thermal occupation in $\ket{g}$
while 10\% and 8\% occupation in $\ket{-}$
and $\ket{+}$, respectively. These values
were inspired from the numerical calculation
of Fig.~5(b) in the main text, which resulted
in a good match with the experiment.
The eigenstates $\ket{\alpha}$, $\ket{\beta}$
and $\ket{\gamma}$ are considered to have zero thermal
occupation due to the high value of their energy.

\subsection{Data recording procedure for instability measurements}

We describe the details of the data gathering routine 
for CEQA and power spectral density (PSD) measurements.
For the CEQA experiment, we use a vector network analyser(VNA) 
to measure the probe transmission, whereas a separate signal 
generator supplies the pump signal. Both microwave units are
synchronized using the 10~MHz reference signal.
Since we use very low probe powers, we record three traces of 
the transmissions, which are later averaged to reduce the 
trace noise.
The measurements are taken at a bandwidth of 10~Hz to improve 
the signal-to-noise.

For the instability measurements, we record the power spectral
density (PSD) of the outgoing microwave signal using a signal
analyzer.
For Fig.~3(a) of the main text, the PSD is recorded at a resolution 
bandwidth(RBW) of 3~Hz and with 200 averages, which takes 2 minutes 
to acquire each data point.
The PSD is recorded around mechanical sideband frequency $\omega_d +\omega_m$ 
with 1~kHz span, where $\omega_d$ is the pump frequency.

For the data corresponding to mechanical instability in Fig.~3(b) of main text,
the PSD is recorded in a span of 30~MHz with center frequency $\omega_d$.
The spectrum analyzer RBW is set to 5~kHz and average to 1000, which take
nearly 20~s for a single realization of pump power and detuning.
For the phase diagram of instability in Fig.~5(a) and (c)  of the main text, 
we record the PSD neighboring two mechanical sidebands $\omega_d \pm
2\omega_m$ successively, with a span of 2~kHz.
An unstable mechanical resonator will cause a large amplitude 
in the PSD around $\omega_d \pm 2\omega_m$.
Thus, it is sufficient to record the data in these two region 
to conclude about the mechanical instability.
We set the spectrum analyzer's RBW to 5~Hz and average to 10. 
It reduces the measurement time to 8 sec to record two successive PSDs.


%

\end{document}